\documentclass[a4paper,usenatbib,useAMS, twocolumn]{mnras}
\usepackage{times,amssymb,amsmath,color,verbatim,graphicx,url,float,bm,ulem}

\usepackage[titletoc]{appendix}

\usepackage{color, xcolor}
\usepackage{natbib}
\usepackage{graphicx}   
\usepackage{amsmath}    
\usepackage{amssymb}    
\usepackage{multicol}        
\usepackage{bm}     
\usepackage{pdflscape}  
\usepackage{hyperref}
\usepackage{multirow}

\newcommand{\dd}{{\mathrm d}}

\renewcommand{\vec}[1]{\boldsymbol{#1}}
\newcommand{\vt}{\vartheta}

\newcommand{\ontop}[2]{
  \renewcommand{\arraystretch}{0.2}
  \begin{array}{c}
  #1 \\ #2
  \end{array}
  \renewcommand{\arraystretch}{1.0}
}
\newcommand{\lsim}{\ontop{<}{\sim}}
\newcommand{\gsim}{\ontop{>}{\sim}}

\newcommand{\Omegam}{\Omega_{\rm m}}

\newcommand{\psfex}{\texttt{PSFEx}}
\newcommand{\sex}{\texttt{SExtractor}}
\newcommand{\swarp}{\texttt{SWarp}}
\newcommand{\scamp}{\texttt{SCAMP}}
\newcommand{\lensfit}{\texttt{LensFit}}
\newcommand{\gaia}{\texttt{GAIA}}
\newcommand{\tmass}{\texttt{2MASS}}
\newcommand{\vsttube}{\texttt{VST-Tube}}

\title[VOICE shear catalogue]{Weak lensing Study in VOICE Survey I: Shear
  Measurement}

\author[Fu et al.]  { Liping Fu$^{1}$\thanks{Corresponding author:
    \texttt{fuliping@shnu.edu.cn}}, 
  Dezi Liu$^{2,1,3}$,
  Mario Radovich$^{4}$,
  Xiangkun Liu$^{2}$, Chuzhong Pan$^{3}$, 
  \newauthor  Zuhui Fan$^{2,3}$, Giovanni Covone$^{5,6,7}$,  Mattia Vaccari$^{8,9}$ ,
 Valeria Amaro$^{5,1}$,
  \newauthor    Massimo Brescia$^{7}$, Massimo Capaccioli$^{5,7}$,
Demetra De Cicco$^{5}$,
 Aniello Grado$^{7}$, 
\newauthor Luca  Limatola$^{7}$,  Lance Miller$^{10}$,
 Nicola R. Napolitano$^{7}$,
Maurizio Paolillo$^{5,6,7}$,
\newauthor Giuliano Pignata$^{11,12}$
  \\\\
  $^1$Shanghai Key Lab for Astrophysics, Shanghai
  Normal University, Shanghai 200234, China\\
  $^2$  South-Western Institute for Astronomy Research, Yunnan University, Kunming 650500, Yunnan, China\\
  $^3$Department of Astronomy, Peking University, Beijing
  100871, China\\
  $^4$ INAF - Osservatorio Astronomico di Padova, via dell‘Osservatorio 5, I-35122 Padova,
  Italy\\
$^{5}$Dipartimento di Fisica ``E. Pancini", Universit\`a degli Studi Federico II, Napoli 80126, Italy \\
$^{6}$INFN, Sezione di Napoli, Napoli 80126, Italy\\
$^{7}$INAF--Osservatorio Astronomico di Capodimonte, Salita Moiariello 16, Napoli 80131, Italy\\
$^{8}$Department of Physics \& Astronomy, University of the Western Cape, Robert Sobukwe Road,
7535 Bellville, Cape Town, South Africa \\
$^{9}$INAF - Istituto di Radioastronomia, via Gobetti 101, 40129 Bologna, Italy\\
$^{10}$Department of Physics, Oxford University, Keble Road, Oxford OX1 3RH, UK\\
$^{11}$Departemento de Ciencias Fisicas, Universidad Andres Bello,
Santiago, Chile\\
$^{12}$ Millennium Institute of Astrophysics (MAS), Nuncio Monseñor Sótero Sanz 100, Providencia, Santiago, Chile
 }

\begin{document}
\maketitle
\begin{abstract}
  The VST Optical Imaging of the CDFS and ES1 Fields (VOICE) Survey is
  a Guaranteed Time program carried out with the ESO/VST telescope to
  provide deep optical imaging over two 4 deg$^2$ patches of the sky
  centred on the CDFS and ES1 pointings.  We present the cosmic shear
  measurement over the 4 deg$^2$ covering the CDFS region in the
  $r$-band using $\lensfit$.  Each of the four tiles of 1 deg$^2$ has
  more than one hundred exposures, of which more than 50 exposures
  passed a series of image quality selection criteria for weak lensing
  study.  The $5\sigma$ limiting magnitude in $r$- band is 26.1 for
  point sources, which is $\gsim1$ mag deeper than other weak lensing
  survey in the literature (e.g. the Kilo Degree Survey, KiDS, at
  VST).  The photometric redshifts are estimated using the VOICE
  $u,g,r,i$ together with near-infrared VIDEO data $Y,J,H,K_s$.  The
  mean redshift of the shear catalogue is 0.87, considering the shear
  weight.  The effective galaxy number density is 16.35
  gal/arcmin$^2$, which is nearly twice the one of KiDS.  The
  performance of $\lensfit$ on such a deep dataset was calibrated
  using VOICE-like mock image simulations. Furthermore, we have
  analyzed the reliability of the shear catalogue by calculating the
  star-galaxy cross-correlations, the tomographic shear correlations
  of two redshift bins and the contaminations of the blended galaxies.
  As a further sanity check, we have constrained cosmological
  parameters by exploring the parameter space with Population Monte
  Carlo sampling.  For a flat $\Lambda$CDM model we have obtained
  $\Sigma_8 = \sigma_8(\Omegam/0.3)^{0.5} = 0.68^{+0.11}_{-0.15}$.

\end{abstract}

\section{Introduction}

Gravitational lensing is the image distortion of background galaxies
(sources) due to the differential deflection of their light caused by
foreground masses (lenses). The induced coherent shape distortion of
source images is referred to as weak lensing shear, and it is
typically much smaller than the intrinsic ellipticity of the source
galaxies. Such signals can only be measured in a statistical way by
averaging over a large sample of galaxies. Weak lensing effects depend
sensitively on the growth of large-scale structures and the expansion
history of the Universe, thus representing a probe complementary to
other observables in order to constrain cosmological models
\citep[e.g.][]{2013ApJS..208...19H, 2016A&A...594A..13P}. Furthermore,
the gravitational nature of weak lensing makes this effect
particularly important in probing the dark side of the Universe
\citep[e.g.][]{2001PhR...340..291B,2014RAA....14.1061F,
  2015RPPh...78h6901K, 2017arXiv171003235M}.

The progresses of cosmological studies based on weak lensing rely on
the developments of wide-field imaging surveys. The
Canada-France-Hawaii Telescope Lensing Survey
\citep[CFHTLenS,][]{CFHTLenS-sys} has shown that cosmic shear is a
powerful cosmological probe \citep{2013MNRAS.430.2200K,
  CFHTLenS-2pt-tomo, CFHTLenS-3pt, 2016PhRvL.117e1101L}. On-going
surveys, such as the Dark Energy Survey
\citep[DES,][]{2016PhRvD..94b2002B, 2016MNRAS.460.2245J}, the
Kilo-Degree Survey \citep[KiDS,][]{ 2015MNRAS.454.3500K,
  2017MNRAS.465.1454H} and the Hyper Suprime-Cam (HSC) survey
\citep{2018PASJ...70S...4A, 2018PASJ...70S..25M} are enlarging the sky
coverage to a few thousands square degrees. In the coming years,
next-generation weak-lensing projects such as the Euclid
mission\footnote{http://sci.esa.int/euclid}, the wide Field Infrared
Survey Telescope (WFIRST\footnote{https://wfirst.gsfc.nasa.gov/}) and
the Large Synoptic Survey Telescope
(LSST\footnote{https://www.lsst.org}) will produce a large
breakthrough in survey volume and depth, making high-precision weak
lensing studies possible.

While the large-sky coverage is essential to minimize the cosmic
variance, the survey depth of weak lensing surveys is crucial to study
the evolution of large-scale structures over the widest redshift
range.  However, deep imaging surveys present different challenges.
The higher number density of background galaxy (few tens to hundred
galaxies per square arcminute) causes crowding problems, making object
de-blending a serious issue, particularly for ground-based
observations.  Moreover, due to the more stringent observing
conditions, deep surveys for weak lensing are more difficult to plan
and carry-out, compared to wide surveys.  Despite that, there are a
number of deep small sized surveys which have set the ground in the
field. CFHTLS Deep \citep{2006A&A...452...51S} has been the first
generation of these deep surveys, and released a 4 deg$^2$ shear
catalogue with the depth of $i = 25.5$.  More recently, the Deep Lens
Survey \citep[DLS,][]{2013ApJ...765...74J,2016ApJ...824...77J}
successfully derived cosmological constraints using a cosmic shear
catalogue with a limit of $r = 27$ mag and a mean source redshift of
$z_s\sim 1$ over 20 deg$^2$.  \cite{2010A&A...516A..63S} presented the
space-based galaxy shape measurements Hubble Space Telescope Cosmic
Evolution Survey (COSMOS) and found evidence of the accelerated
expansion of the Universe from weak lensing tomography. This result
has been obtained with data collected over a field of view of only
1.64 deg$^2$, but with a very high galaxy number density, 76
arcmin$^{-2}$ with limiting magnitude $i < 26.7$ mag.

The VLT Survey Telescope (VST) Optical Imaging of CDFS and ES1
\citep[VOICE, co-PIs: Giovanni Covone \& Mattia
  Vaccari,][]{2016heas.confE..26V} is a Guaranteed Time of Observation
(GTO) survey preformed with the ESO/VST telescope (\citealt{VST})
operating on Cerro Paranal (Chile).  VOICE shared observations with
the SUpernova Diversity And Rate Evolution (SUDARE), another VST GTO
survey, to cover the CDFS sky region \citep{2015A&A...584A..62C,
  2017A&A...598A..50B}.  SUDARE has observed the common fields in the
$g,r,i$, optimizing the strategy in order to search and characterize
supernovae at intermediate redshift ($0.3 \lsim z \lsim 0.6$).  The
VOICE team has been in charge of the $u$ band observations of the same
area.  For their science case, SUDARE required less stringent
constraints on image quality, however the number of epochs was so
large that the total amount of data with image quality within VOICE
specs in $g,r,i$ allowed us to reach the necessary depth in the
stacked images required by the VOICE science objectives, including
weak lensing.

The two selected fields, VOICE-CDFS and VOICE-ES1, have been also
observed by other facilities on a wide wavelength range, including
GALEX (UV), VISTA-VIDEO (NIR), Spitzer-SERVS (MIR), Herschel-HerME
(FIR), Spitzer SWIRE (IR), and ATLAS (radio).  Adding optical data
from VOICE has made these fields extremely valuable for a large range
of astrophysical studies. One of the science drivers for VOICE is to
detect clusters of galaxies at relatively high redshifts, and to study
their mass distributions using weak lensing signals of galaxies in the
fields.

The VOICE survey uses the same telescope, detector (OmegaCAM) and
optical filters as KiDS. The $r$-band data are used for weak lensing
measurements. Differently from KiDS, where each pointing is observed
only in one epoch consisting of five consecutive exposures, the VOICE
survey holds multiple-epoch observations for each pointing of the
$r$-band with total number of exposures over a hundred. For the data
used for weak lensing shear measurements, the $5\sigma$ limiting
magnitude for point source in $r$-band co-added images reaches
$r=26.1$ mag within 2\arcsec aperture diameter, which is about $1.2$
magnitude deeper than KiDS data.

 As in KiDS \citep[][hereafter K15]{2015MNRAS.454.3500K}, we used
 $\lensfit$
 \citep{2007MNRAS.382..315M,2008MNRAS.390..149K,2013MNRAS.429.2858M}
 to measure the galaxy shapes.  To this end, some preliminary steps
 were required.  First, the observing conditions varied significantly
 from epoch to epoch and we needed to go through a severe quality
 control of the individual exposures. Second, we needed to adapt the
 $\lensfit$ parameters for our dataset, since VOICE data are deeper
 than CFHTLenS and KiDS \citep{2017A&A...604A.134D}. To validate the
 setup and calibrate the shear measurement, we made use of dedicated
 simulations which have been presented in a companion paper
 \cite{2017LiuDZ}.

The structure of this paper is organized as follows. In
Section~\ref{sec:survey}, we describe VOICE data and data
reduction. The shape measurement procedures, the calibration from
VOICE-like simulation and the photometric redshift are presented in
Section~\ref{sec:shear-cat}. Two-point correlation analyses and null
tests for shear systematics are presented in
Section~\ref{sect:shear}. To further demonstrate the quality of our
shear measurements, in Section~\ref{sec:cosmo}, we show the
cosmological constraints of $\Omega_{\rm m}$ and $\sigma_8$ derived
from cosmic shear two-point correlations. The summary is given in
Section~\ref{sec:summary}.

\section{The survey}
\label{sec:survey}

This paper focuses on the VOICE-CDFS field, which covers about 4.9
deg$^2$.  It is composed by four {\it tiles} (CDFS1, CDFS2, CDFS3,
CDFS4), about 1 deg$^2$ each. The pixel scale of the OmegaCAM CCDs is
$0.21''$. The center of the VOICE-CDFS field is RA$=03^{\rm h}32^{\rm
  m}30^{\rm s}$ and DEC$=-27^{o}48\arcmin30\arcsec$.  The observations
started in October 2011, and ended in 2015. Each tile was observed in
four optical bands $u,g,r,i$ with exposure time of 600s ($u$), 360s
($g$ and $r$) and 400s ($i$), respectively.  The $r$-band data were
used, in addition to the weak lensing study presented here, for
variability based search of supernovae \citep{2017A&A...598A..50B} and
Active Galactic Nuclei \citep{2015A&A...579A.115F,
  2015A&A...574A.112D}.  For each tile, more than one hundred
exposures were taken in the $r$-band.  As in KiDS, a single {\it
  epoch} consists of five consecutive exposures obtained with a
diagonal dithering pattern to cover the detector gaps. The initial
position and the dithering pattern is repeated at any epoch. The
cumulative exposure time ranges from 15.3 to 20.9 hours for the four
fields. The total exposure time for the other three bands is shorter
as shown in Table~\ref{CDFS}.  As $\sim 100$ VOICE exposures are
distributed over four years, the image quality and the point spread
function (PSF) of the individual exposures varies significantly in
exposures from different epochs.
 \begin{table}
        \caption{
          The total exposure time  (in hours)
          of four VOICE-CDFS fields in the $u,g,r,i$ bands before
          applying any image quality selections (Sect.~\ref{sec:reduction}).}
        \centering
        \label{CDFS}
        \begin{tabular}{c|l|l|l|l|}
         \hline

            \cline{1-5}       & $u$ & $g$ & $r$ & $i$  \\
            \hline
              CDFS1     & 5.20
                    & 5.64 & 20.90 & 8.41\\ 
              CDFS2      & 6.50
                    & 4.83 & 15.30 & 4.38\\ 
              CDFS3     & 0.83 & 6.94 & 20.60 & 9.47\\ 
              CDFS4      & 0.83 & 5.43 & 18.50 & 8.51\\ 
           \hline      
 
        \end{tabular}
  \end{table}

\subsection{Exposure selections}
\label{sec:reduction}

The data reduction was performed using the pipeline $\vsttube$
\citep{2012MSAIS..19..362G}. As described in detail in
\cite{2015A&A...584A..62C}, $\vsttube$ performs over scan correction,
flat fielding, CCD gain harmonization, illumination correction, and
cosmic ray removal.

 Since the shear signal is very weak, about an order of magnitude
 smaller than the intrinsic ellipticity of galaxies, we have applied
 very strict image selection criteria.  VOICE $r$-band observations
 were carried out over 4 years, therefore, the observing conditions
 show significant variations among epochs.  In order to obtain an
 homogeous dataset and maximize the quality of our shear measurements,
 we have filtered our data according to seeing and its variations
 within the field of view before further data processing (i.e., image
 co-adding, object detection and shape measurements).

\begin{figure}

 \resizebox{0.9\hsize}{!}{
   \includegraphics[bb = -14 1 445 380]{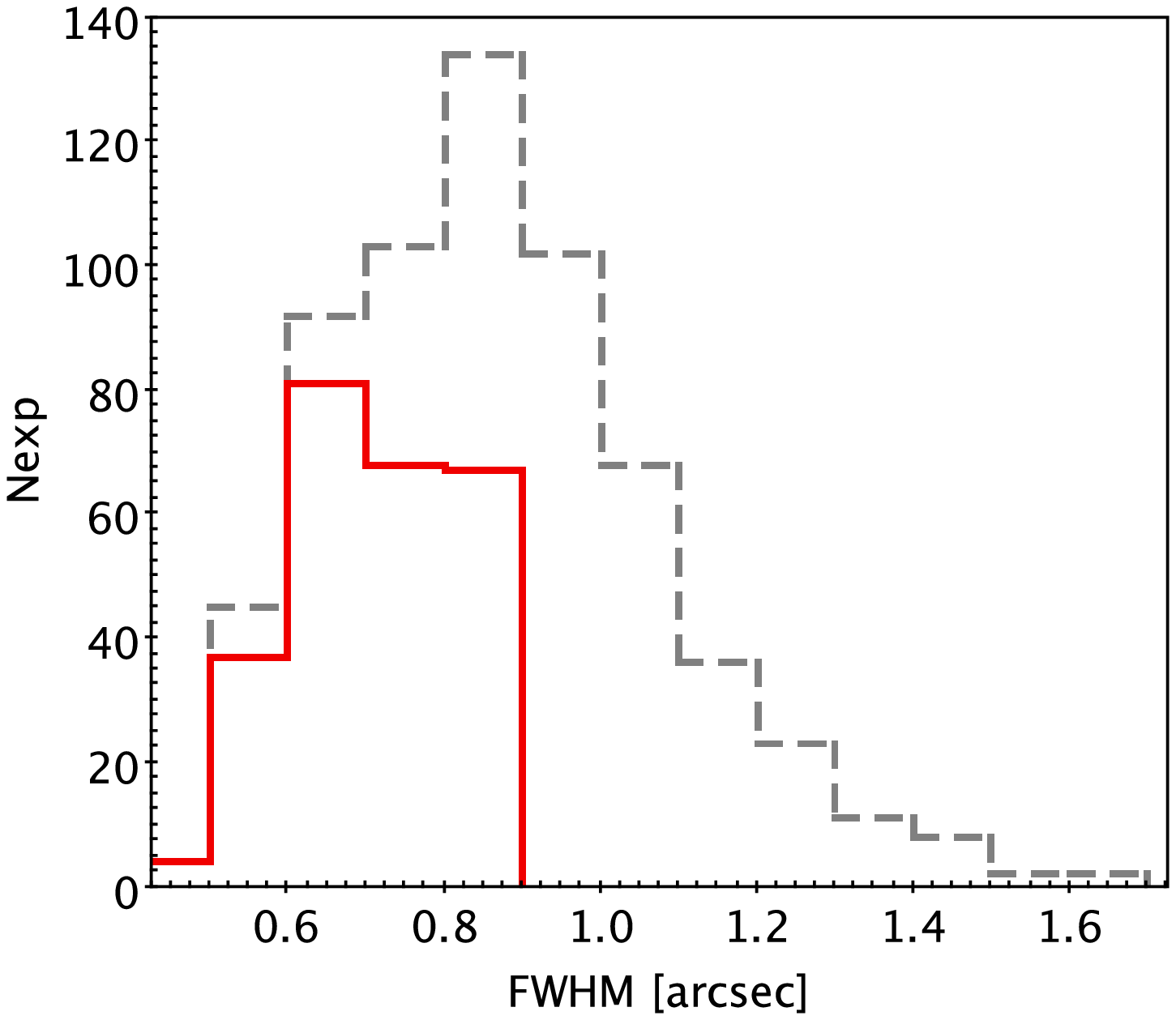}
   }
  \resizebox{0.97\hsize}{!}{
    \includegraphics[bb = -18 1 470 380]{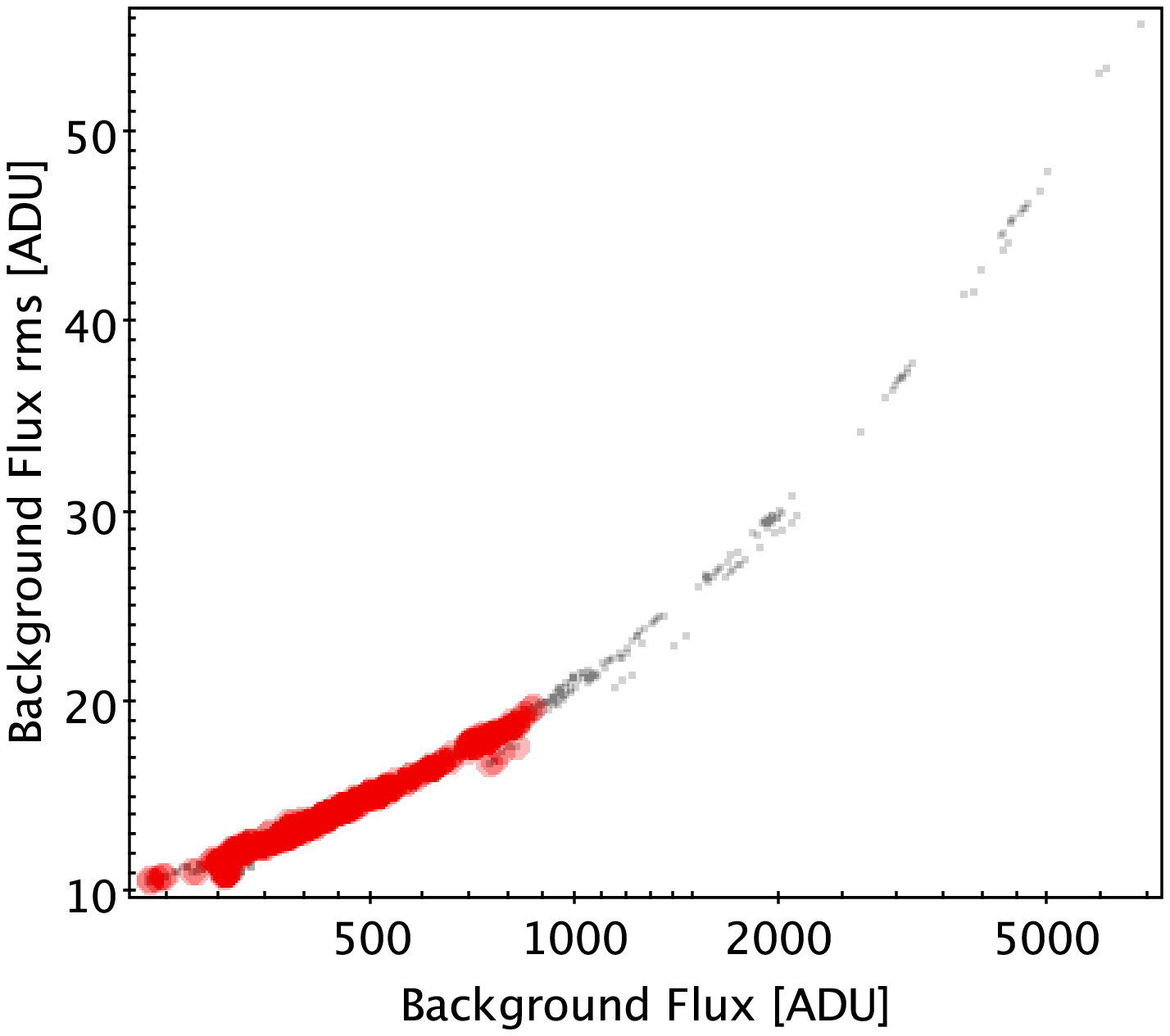}
  }

  \caption{The  PSF FWHM distribution ({\it top}) and the correlation
    between 
    background value and its CCD to CCD dispersion ({\it bottom}) of all
    $r$-band exposures (grey). The final selected exposures
    are shown in red. }
   \label{fig:seeing}
\end{figure}

The PSF full width at half maximum (FWHM) of different exposures in
the $r$-band ranges from 0.4$\arcsec$ to 1.5$\arcsec$ as shown in the
top panel of Fig.~\ref{fig:seeing}. The median value is
0.86$\arcsec$. Weak lensing studies focus on background galaxies which
are mostly faint and small.  Because seeing smears galaxy images if
they are significantly smaller than the seeing disc, we have selected
only those exposures with seeing smaller than $0.9\arcsec$.

The sky background brightness can also affect object detection and
shape measurement. The background values calculated by $\sex$
\citep{2011ASPC..442..435B} spread in an extremely wide range, from a
few hundreds to a few thousands Analog-to-Digital Units (ADUs). We
assigned the median background value of the 32 CCDs as the reference
background flux value of each exposure. As shown in the bottom panel
of Fig.~\ref{fig:seeing}, the rms value is strongly correlated with
the background flux.  Most of the exposures showed relatively small
background flux and small variations from CCD to CCD.  We have then
applied a cut on the background rms dispersion in order to have a
homogeneous background noise. After several iterations examining the
B-mode in the shear two-point correlations, the exposures with
background rms dispersion over 20 were rejected in the shear analysis,
corresponding to a background flux cut of $< \sim$900 ADU.

  In order to have a uniform depth from epoch to epoch, we further
  reviewed the remaining exposures, and only kept those epochs with at
  least four exposures passing the selection criteria. In conclusion,
  about one-third of total exposures were used for weak lensing
  analysis, as shown in red in Fig.~\ref{fig:seeing}.  The number of
  useful exposures for the four tiles is 62, 54, 79 and 62,
  respectively.  The final mosaic reaches a 5$\sigma$ limiting
  magnitude of $r_{\rm AB} = 26.1$ within 2$\arcsec$ aperture diameter
  for point sources. The average limiting magnitude for $u,g,i$ bands
  is 25.3, 26.4, 25.2, respectively.

\subsection{Astrometric calibration}
\label{sec:Acalibration}

\begin{figure}

  \resizebox{0.9\hsize}{!}{
    \includegraphics[bb = 1 1 430 400]{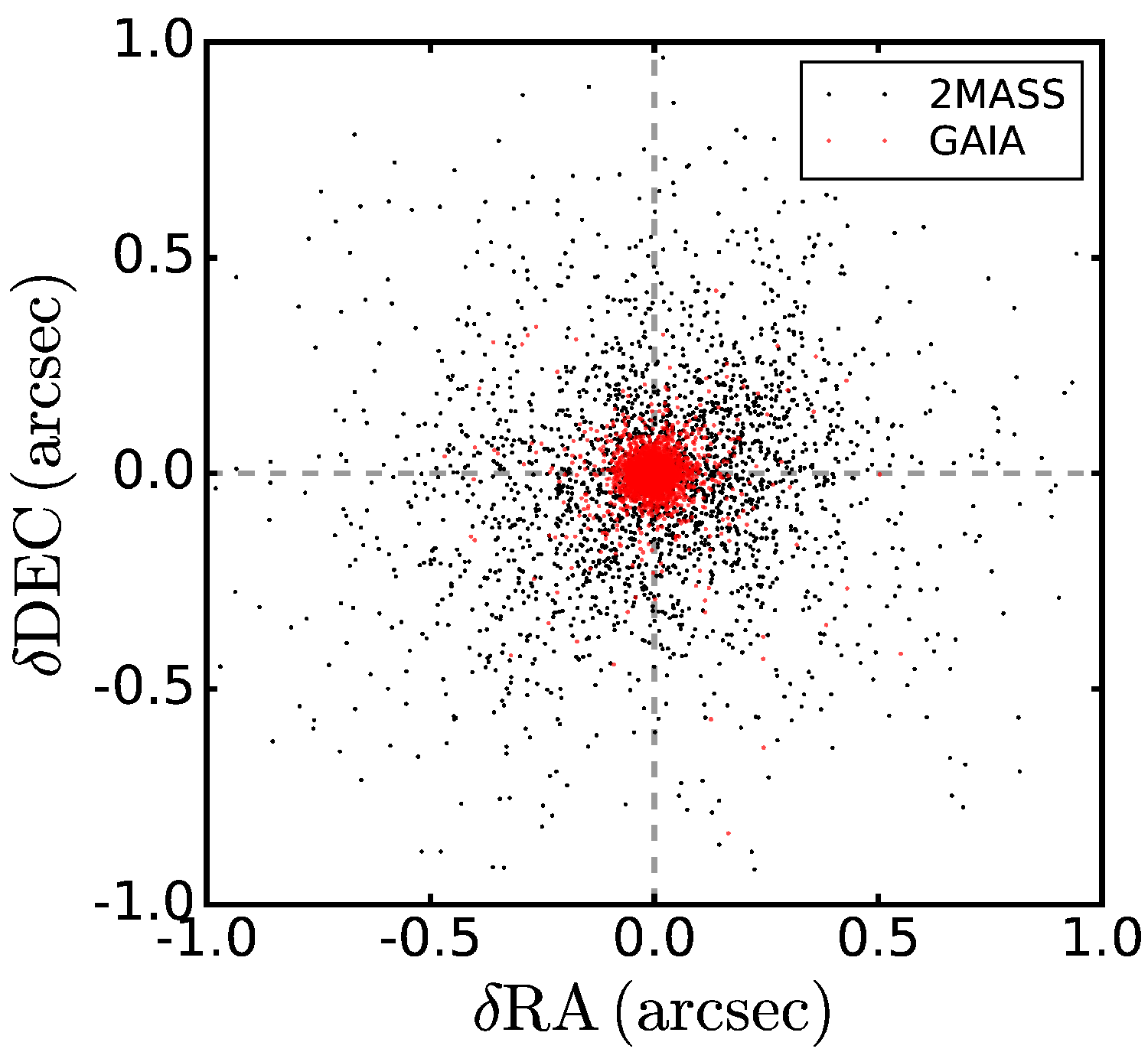}
  }
  
  \caption{The RA and Dec difference  of matched objects between VOICE and
    $\tmass$ (black), or VOICE and $\gaia$ (red).    }
   \label{fig:gaia}
\end{figure}

  The astrometric calibration of each tile has been performed
  separately using the software
  $\scamp$\footnote{https://www.astromatic.net/2010/04/20/scamp-1-7-0-release}.
  Only exposures that passed our selection criteria were used
  simultaneously for the calibration in order to improve the internal
  accuracy. The external accuracy depends on the choice of reference
  catalogue. We performed two sets of calibrations using $\tmass$
  \citep{2006AJ....131.1163S} and $\gaia$ \citep{2016A&A...595A...2G},
  respectively. The calibrated exposures were co-added by
  $\swarp$\footnote{https://www.astromatic.net/2010/09/04/swarp-2-19-1-release}
  to produce the final stacked image used for source detection.  We
  have matched the objects between the reference catalogue and the
  VOICE deep image:  the matched objects are 6634 and 10555 for
    $\tmass$ and $\gaia$, respectively. As shown in
  Fig.~\ref{fig:gaia}, the astrometric dispersion based on $\gaia$
  (0.056$\arcsec$) is about four times smaller than that from $\tmass$
  (0.19$\arcsec$),  since $\gaia$ has smaller intrinsic
    astrometric uncertainties and a higher matched number of stars
    with respect to $\tmass$.  Therefore, we have chosen $\gaia$ as
  the absolute reference for the VOICE astrometric calibration.
  
\begin{figure}
  \resizebox{0.9\hsize}{!}{
    \includegraphics[bb = 1 1 460 500]{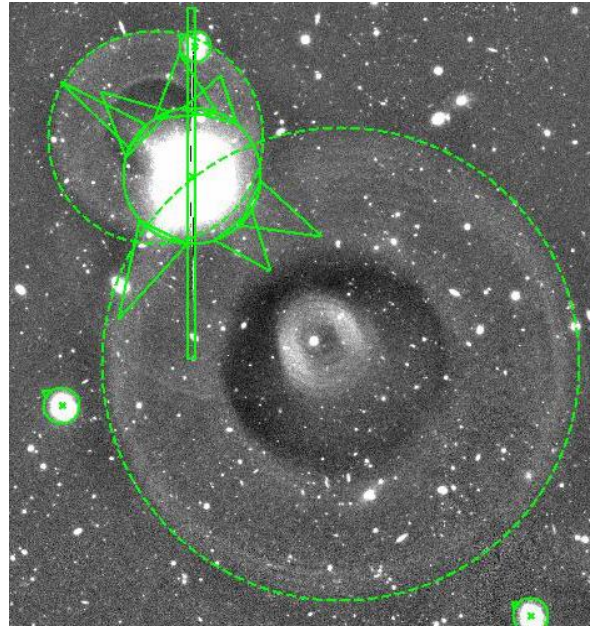}
  }
    \caption{Example of Masked regions covering  saturated stars,
      halos, spikes and the other defects in the CDFS2.
     }
   \label{fig:mask}
\end{figure}

  \begin{table}
    \caption{ The number of sources used in our analysis in the four
      CDFS tiles in the $r$-band: $\rm N_{star}$ is the number of
      stars used for PSF correction; $\rm N_{\rm gal}$ is the number
      of galaxies detected from the co-added deep image; $\rm N_{\rm
        shear}$ is the number of galaxies with $\lensfit$ non-zero
      weight; $\rm N_{\rm exclude}$ is the number of galaxies excluded 
      before model fitting; $\rm N_{\rm wzero}$ is the number of
      galaxies that passed exclusion selection but failed in
      $\lensfit$ model fitting with zero weight. }
    \label{Ngal}
    \centering
        \begin{tabular}{l|l|l|l|l|}
         \hline
         \hline
             & CDFS1 & CDFS2 & CDFS3 & CDFS4 \\
            \hline
       
            $\rm N_{star}$ & 2878 & 2807 & 2851 & 2774 \\
           $\rm N_{\rm gal}$     & 129505 & 125032 & 126360 & 125295 \\
           $\rm N_{\rm shear}$   & 84406 & 83425 & 78445 & 77499 \\
           \hline                         
           $\rm N_{\rm exclude}$  & 24686 & 22946 & 25830 & 23914 \\
           $\rm N_{\rm wzero}$    & 20413 & 18661 & 22085 & 23882 \\
           
           \hline      
 
        \end{tabular}
  \end{table}

\subsection{Mask}
\label{sec:mask}

Saturated stars and their surrounding areas have to be masked because
the flux measured in those regions can be affected by strong
systematic errors. Those areas were identified by the automatic mask
software Pullecenella \citep{Huang2011, 2015A&A...582A..62D}, which
has been created specifically to treat the VST images.  For
$\lensfit$, the galaxy model fitting is performed on each individual
exposure.  Thus the masks were not produced from the deep co-added
images in order to avoid over masking. Instead, we masked the affected
areas of the individual epochs, i.e., the stacked images over five
consecutive and dithered exposures.  Fig.~\ref{fig:mask} shows an
example of masked regions near saturated stars with a large reflection
halo.  The remaining unaffected area after masking is $\sim$ 84\% of
the original 4.9 deg$^2$ VOICE-CDFS area.

\subsection{Photometric redshift  catalogue description}
\label{sec:zcat}

For each tile all the high-quality, astrometric calibrated exposures
were co-added using $\swarp$ to produce the deep stacked image. Source
positions and star-galaxy classification were performed on the stacked
image.  The $\sex$ software \citep{Bertin1996} was run to generate the
final source catalogue.  The star-galaxy classification was done in
the magnitude-size diagram \citep{Huang2011}, where magnitude and size
are represented by the $\sex$ parameters MAG\_AUTO and
MU\_MAG$-$MAG\_AUTO. Sources with size smaller than the stellar one
were defined as spurious and removed from the catalogue.  As shown in
Table~\ref{Ngal}, about 2800 stars were selected from each tile and
used to measure the PSF. More than $1.25\times 10^5$ galaxies per tile
were selected. This galaxy catalogue was used for the photometric
redshift estimates (photo-$z$) and also as input to the shape
measurement software $\lensfit$ \citep{2007MNRAS.382..315M,
  2008MNRAS.390..149K,2013MNRAS.429.2858M}.

   For photo-$z$ measurements, we employed the optical observations in
   $u, g, r, i$ from VOICE, and the near-infrared $Y,J,H,K_s$ data
   obtained by the VIDEO survey \citep{Jarvis2013} performed with the
   VISTA telescope. The NIR bands cover $>$ 80\% of the VOICE images.
   We did not include the VIDEO $Z$ band since it covers a negligible
   fraction ($<$ 50\%) of the VOICE area. The VOICE and VIDEO stacks
   were produced selecting exposures with a similar cut in the seeing
   ($\le$ 1.0 arcsec). We therefore decided to base our photometric
   redshift estimate on magnitudes measured on apertures of the same
   size in all bands. To this end we used the \texttt{SEP} Python
   library \citep{Barbary2016}: the \texttt{SEP} library implements
   algorithms from the $\sex$ software \citep{Bertin1996} as
   stand-alone functions and classes. We used it to measure
   $u,g,r,i,Y,J,H,K_s$ aperture magnitudes (6$\arcsec$ diameters)
   centered on the source positions in the $r$-band catalogue.
   Compared to the so-called {\it dual-mode} in $\sex$, the
   \texttt{SEP} library allows to perform a list-driven photometry on
   images with different size, scale or center: WCS coordinates from
   the catalogue were converted to pixel positions in the image using
   functions available in the astropy python library and then passed
   to \texttt{SEP}. Background subtraction is also available within
   \texttt{SEP}.

The next step was the removal of residual errors in the calibration of
the photometric zero point. To this end, we benefit from the overlap
of the CDFS fields with the APASS
survey\footnote{\url{https://www.aavso.org/apass}}. We matched $\sim
200$ unsaturated stars (15 $< r < $16 ) in the $gri$. Non-negligible
offsets ($< 0.1$ mag) were found in $g$ (CDFS3 and CDSF4) and $i$
(CDFS3).

Photo-$z$ were finally derived using the {\texttt{BPZ}} software
\citep{Benitez2010}: {\texttt{BPZ}} adopts a Bayesian approach, where
the likelihood that a template fits the colours of a galaxy at a given
redshift is combined with a prior defining the probability to find a
galaxy of that type, as a function of magnitude and redshift. This
allows to reject those solutions which would maximize the likelihood,
but that would be unphysical according the known prior
distributions. The {\texttt{BPZ}} library consists \citep{Benitez2004}
of four modified Coleman, Wu and Weedman types \citep{Coleman1980},
and two Kinney, Calzetti \& Bohlin \citep{Kinney1996} starburst galaxy
templates. The derived photo-$z$ are discussed in
Sect.~\ref{sect:zdis}.

\begin{figure*}
  \resizebox{0.9\hsize}{!}{
    \includegraphics[ angle = -90]{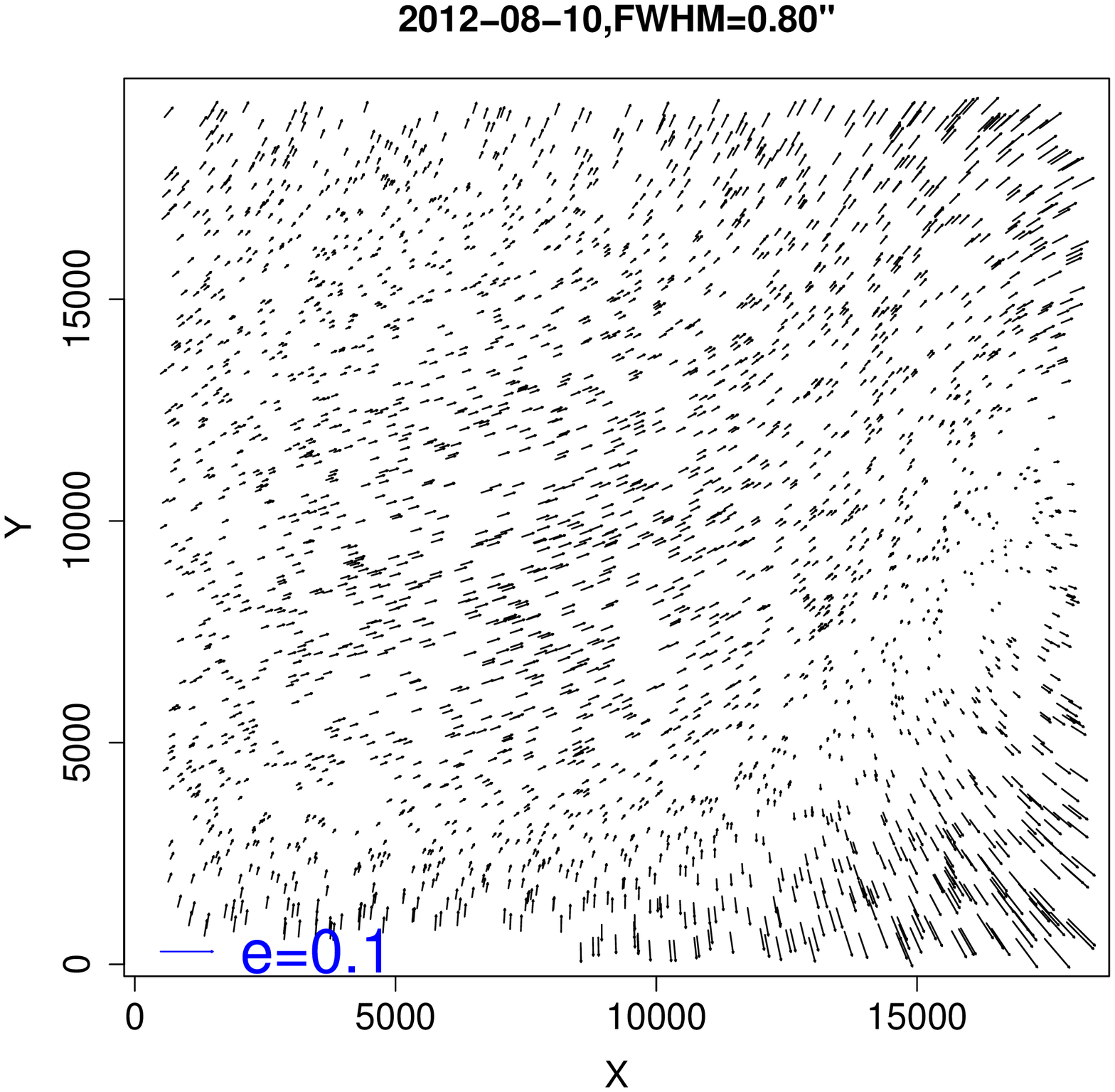}
    \includegraphics[ angle = -90]{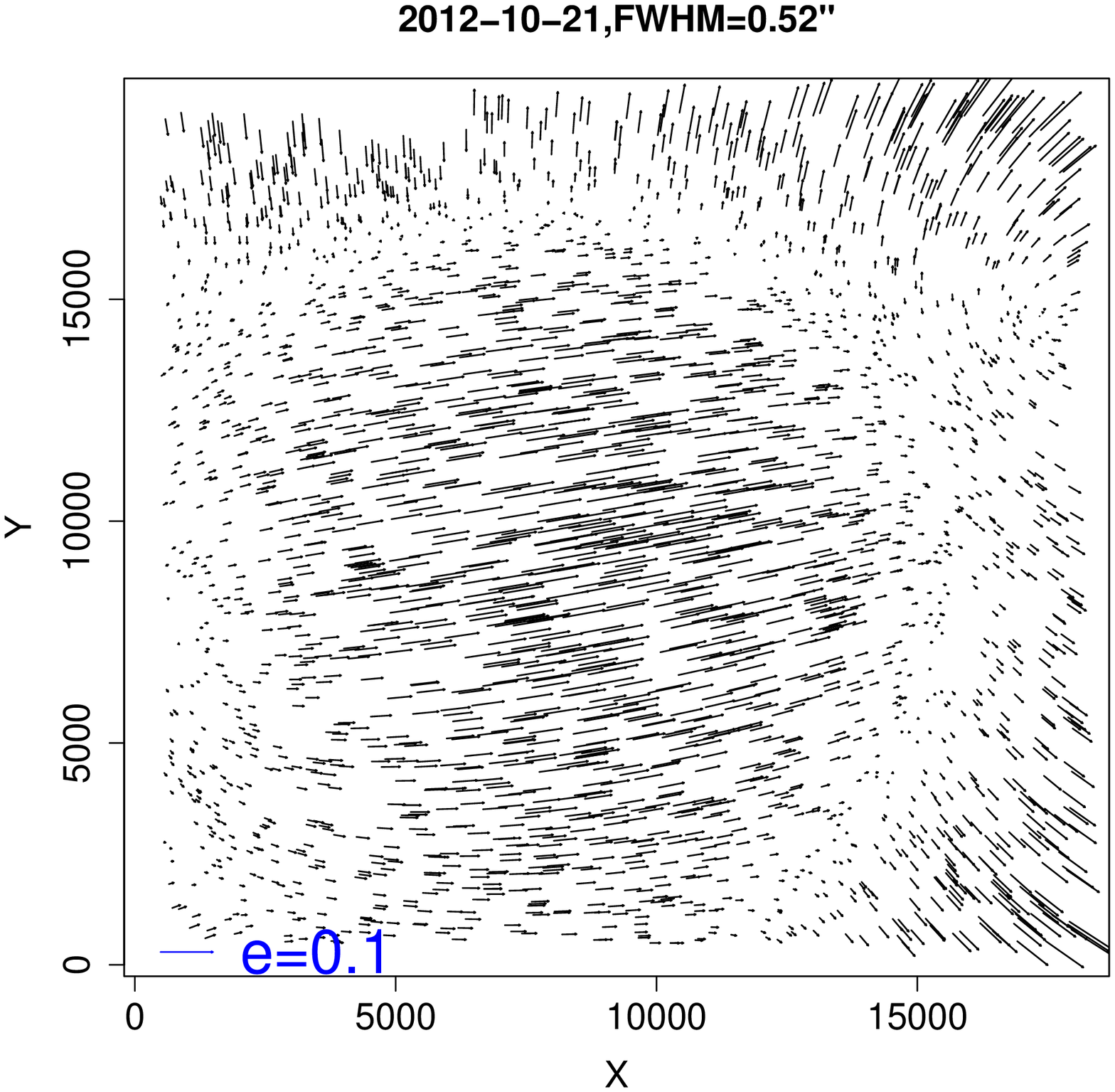}
  }
 \resizebox{0.9\hsize}{!}{
    \includegraphics[ angle = -90]{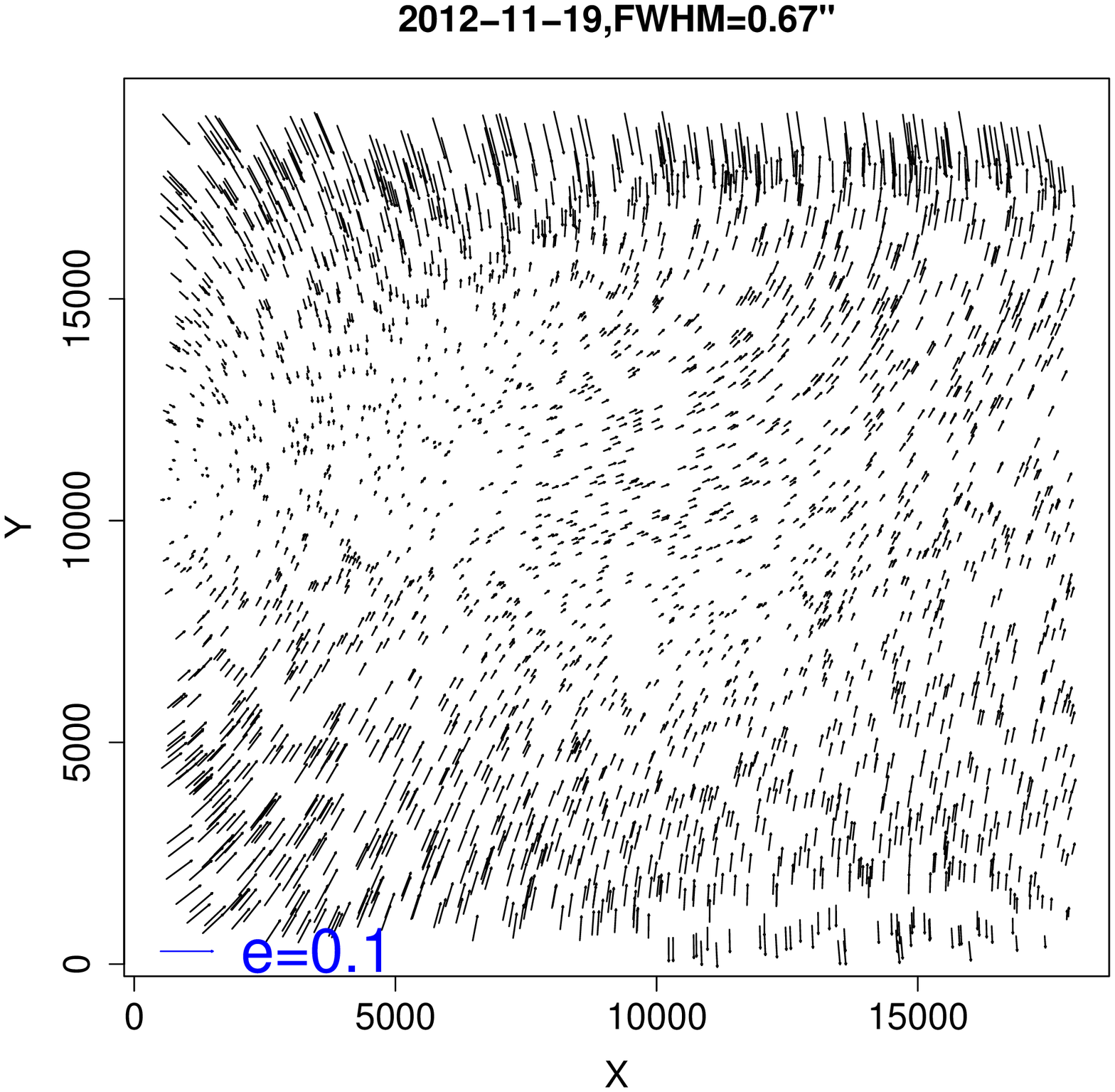}
    \includegraphics[ angle = -90]{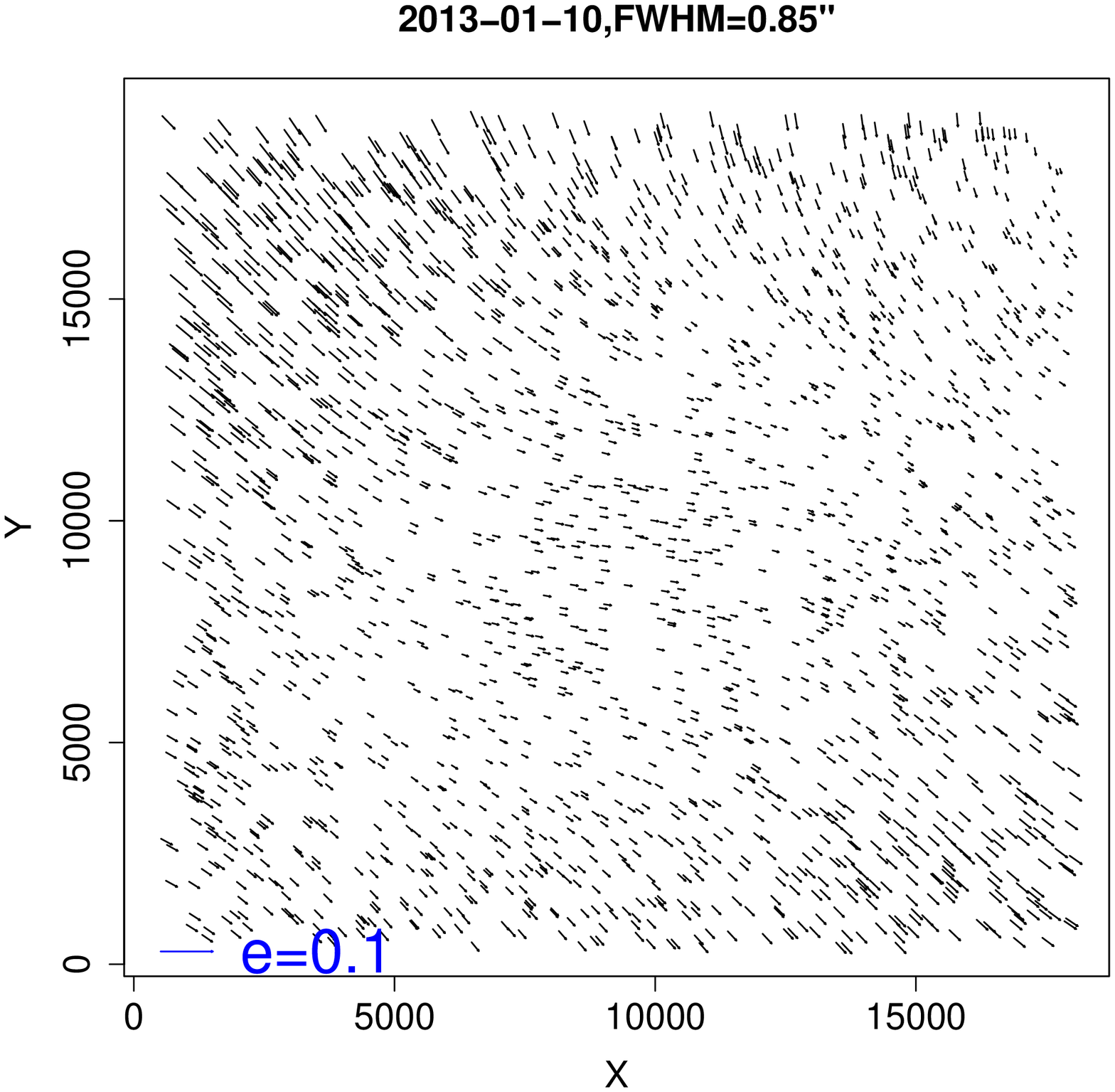}
  }

  \caption{Examples of variations in  PSF patterns in VOICE-CDFS1 for four
    epochs observed from
    summer to winter.  }
   \label{fig:psf}
\end{figure*}

\section{$\lensfit$ Shape measurement}
\label{sec:shear-cat}
The shear measurement accuracy depends sensitively on the data quality
and on the data processing steps, such as the observing conditions,
the quality of the camera, the PSF shape and stability, the background
noise, etc..  It is also crucial to use a reliable shape measurement
algorithm optimized for the considered survey. Image simulations
specifically made for the survey are normally needed to validate the
optimizations and also to quantify the possible biases in the shear
measurements.

KiDS data analyses \citep[e.g.,][]{2017MNRAS.465.1454H} proved that
$\lensfit$ \citep[][hereafter M13]{2013MNRAS.429.2858M} is a suitable
shape measurement algorithm for OmegaCAM images, with an accuracy
reaching $\sim$1\%.

We therefore also adopted $\lensfit$ for the shape
measurement. $\lensfit$ constructs a seven-parameter galaxy model fit
including the galaxy position, flux, scale-length, bulge-to-disc
ratio, and galaxy ellipticity. 
Although the signal-to-noise ratio of
an individual galaxy detected from co-added image is high, 
 using the
co-added image is problematic for high-precision galaxy shape measurement,  
mainly because the co-addition of PSFs of different
shapes and orientations from different exposures may result in a complex
stacked PSF.  
Furthermore, the co-adding procedures (particularly the
interpolation of individual exposures to a common pixel grid)
introduces noise correlation between pixels, which can affect the
shape measurement. 
Thus in $\lensfit$, the model fitting is done on
individual exposures, and the probabilities of the parameters derived
from different exposures for a galaxy are statistically combined to
derive its final shape measurement.  The details of
$\lensfit$  algorithm  are  described in   \citet{2007MNRAS.382..315M,
  2008MNRAS.390..149K} and M13. In the following, we describe the key
issues particularly relevant to the VOICE data.

\subsection{PSF fitting}
\label{sec:psf}

The VOICE observational campaign was distributed over several years.
The PSF patterns of the same tile were very different from month to
month, even night to night.  We show in Fig.~\ref{fig:psf} a few
examples of PSF ellipticity patterns at different epochs in the CDFS1
tile constructed by co-adding PSFs from five exposures within an
epoch. The four epochs were observed at different times, from summer
to winter. Strong temporal variations of PSF are clearly seen.
Furthermore, any sub-optimal optical configuration of the telescope
contributes significantly to the PSF. As discussed in K15, any primary
mirror astigmatism of the curved focal plane of the VST results in an
increasing ellipticity in the center of the field (top-right panel of
Fig.~\ref{fig:psf}), while a tilt of the secondary mirror causes the
increase of ellipticity near one edge of the field (bottom-left panel
of Fig.~\ref{fig:psf}).

Therefore the PSF model fitting is made for each single exposure.
Nevertheless, as shown in Fig.~\ref{fig:psf}, the PSF varies not only
over the full field of OmegaCAM, but also from CCD to CCD. Thus, two
different polynomial fitting models were applied: a 4th order
polynomial fit for the full field-of-view and a 1st order
chip-dependent polynomial for individual CCDs, as done by K15 for the
KiDS survey.

\subsection{Exclusion of galaxies}
\label{sec:blender}
 $\lensfit$ fits each single galaxy in a postage stamp with a size of
48 $\times$ 48 pixels, which is a compromise between a stamp large enough to
obtain a correct model fit, and a stamp small enough for  
fast  processing and fitting. 
 The center of the postage stamp was chosen to be the position
of the galaxy detected from the deep co-added image.  Before the model
fitting, $\lensfit$ performs a few quality checks.  We give a short
summary here, and refer to M13 for more details about the fitting
algorithm.
\begin{enumerate}
\item Galaxies  larger than  the size of the postage
  stamp  were excluded from the  analysis.\\
  
\item To deblend the neighboring galaxies, if more than one object  is
  found within the same postage stamp, the algorithm checks whether
  the neighbour galaxy can be masked by replacing the pixel values of
  the background without contaminating the isophotes of the target
  galaxy.  Comparing the Gaussian-smoothed isophotes of the neighbour
  galaxy measured from the co-added image to the smoothed pixel noise, if
  the signal-to-noise ratio is larger than a defined threshold,  the
  neighbour galaxy will be masked out. Since VOICE is deeper than CFHTLenS
  and KiDS, in order to retain enough  galaxies while still
  suppressing most of the neighbour contaminations, we optimized this
  threshold from two (M13 for CFHTLenS) to five. Imaging simulations
  of \cite{2017LiuDZ} show that this choice does not introduce
  significant bias to the VOICE shear measurements.  More details are
  discussed in Sect.~\ref{sec:simulation} and \cite{2017LiuDZ}.\\

\item If masked pixels are outside the target galaxy's isophote on single
  exposure, the pixels are replaced by the background values and the process
  continues.  If the masked pixels are within the isophote, then that
  exposure will not be used in the joint analysis. \\

\item If the weighted centroid of a galaxy is more than 4 pixels
   away from its stamp center, it implies that there may be
  blended objects existed within the stamp. Thus this galaxy
  is excluded as well.
\end{enumerate}

As shown in Table~\ref{Ngal} (see quantity  ${\rm N}_{\rm exclude}$), the fraction of excluded galaxies from
the above criteria is about 19\%.

\subsection{Shear catalogue}
\label{sec:shear}

 %
\begin{figure}
\resizebox{0.9\hsize}{!}{
        \includegraphics[bb = 145 271 426 520]{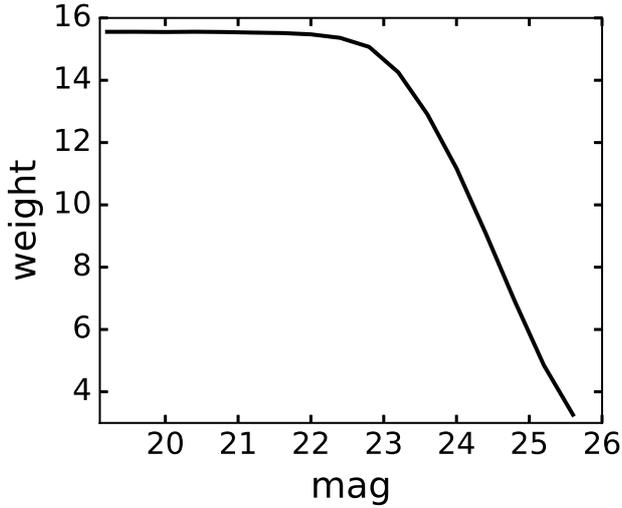}
}
  \caption{Shear  averaged weight as the function of  the $r$-band galaxy magnitude.}
   \label{fig:wmag}
\end{figure}

\begin{figure}
\resizebox{0.9\hsize}{!}{
        \includegraphics[bb = -10 1 470 400]{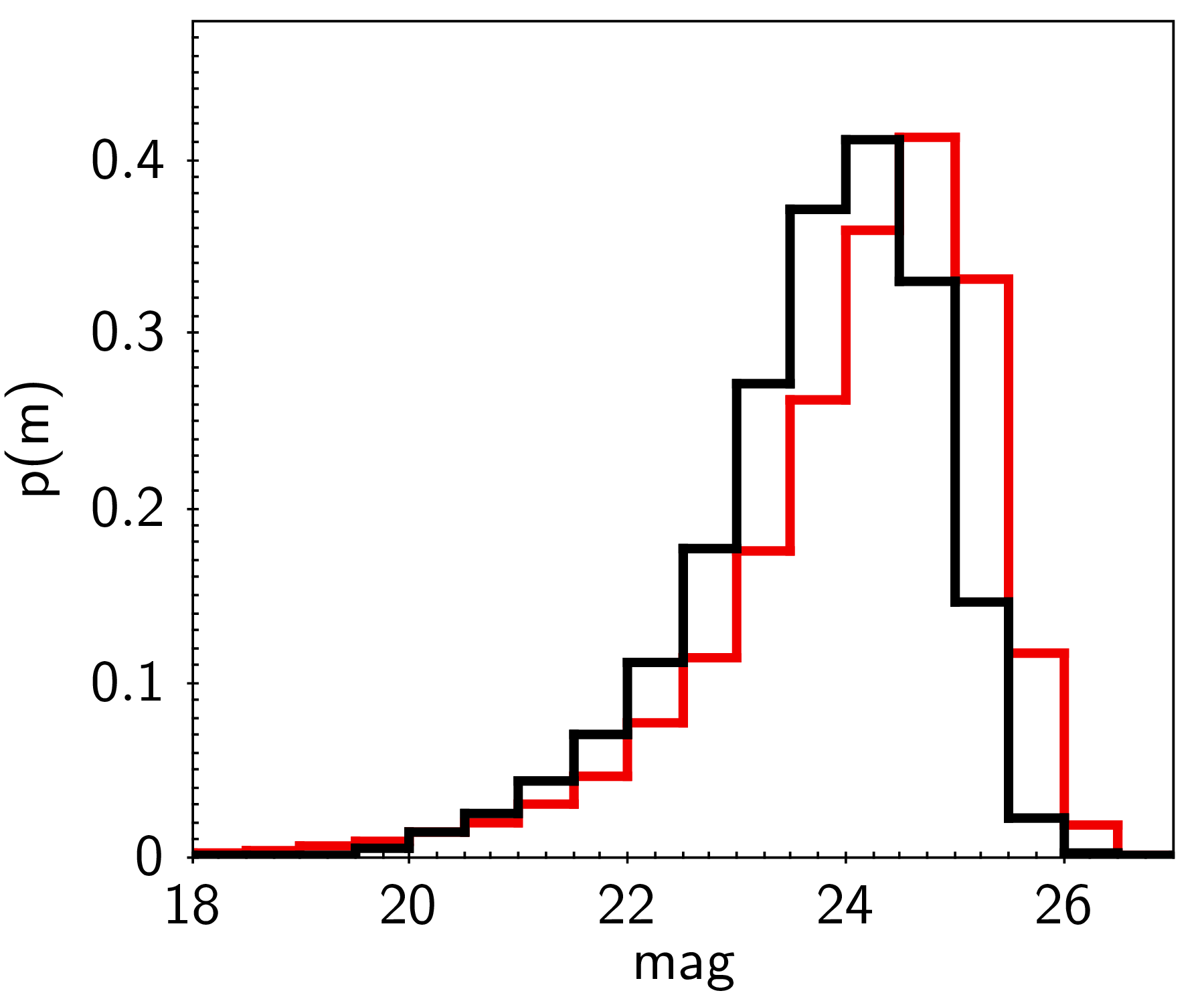}
}
  \caption{The normalized magnitude distribution of galaxies in the four
    CDFS fields without (red) and with (black) shear weight. }
   \label{fig:maghis}
\end{figure}

 $\lensfit$ defines the galaxy weight taking into account both the
shape-noise variance and ellipticity measurement-noise variance
(M13). About 17\% of total galaxies failed in galaxy model fitting
although they passed the exclusion selection. They were given a weight
of zero, and their numbers are shown as ${\rm N}_{\rm wzero}$ in
Table~\ref{Ngal}.  As faint galaxies are much noisier than bright
ones, their weights are much lower as shown in
Fig.~\ref{fig:wmag}. The magnitude distribution of the non-zero weight
galaxies is shown in Fig.~\ref{fig:maghis}. The peak magnitude of the
weighted distribution is about $24.2$ mag, which is about $1.0$ mag
deeper than the $\lensfit$ selected galaxies in KiDS.

 In order to have continuous coverage of CDFS fields, an overlap of
 3$\times$7 arcmin$^2$ has been taken among the four tiles.  Thus
 galaxies from the overlapping regions have to be dealt with
 separately, if they are detected more than once.  Due to astrometric
 errors, some galaxy positions may be slightly different in the
 overlap region of different exposures. If a pair of galaxies has a
 separation of less than $3$ pixels, we considered them as a single
 galaxy and only kept the higher signal-to-noise measurement result.

The final shear catalogue has over $3\times 10^5$ galaxies with
non-zero weight, corresponding  to  an effective weighted galaxy number
density 16.35 arcmin$^{-2}$, which is about double of the density in
the KiDS survey.

\subsection{The photometric redshift distribution}
\label{sect:zdis}
\begin{figure}

  \resizebox{0.9\hsize}{!}{
    \includegraphics[bb = -10 1 420 390]{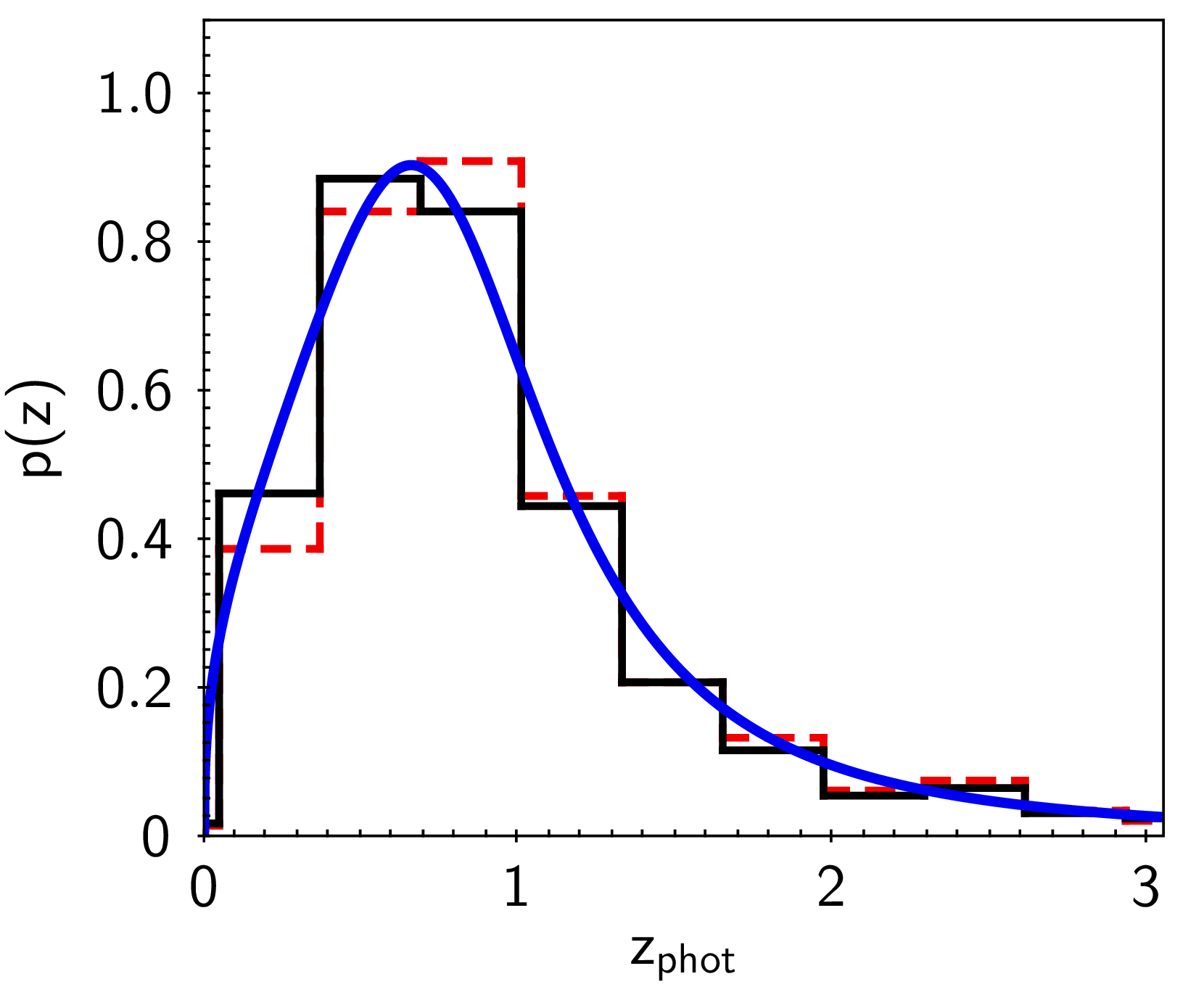}
  }
  
  \caption{The normalized distribution of photo-$z$ (peak value of PDF)
    of VOICE galaxies without (red dash line) and with (black solid
    line) shear weight. The solid blue curve is the best fit of
    photo-$z$ with weight.  }
   \label{fig:photoz}
\end{figure}

The shear catalogue was matched to the photo-$z$ catalogue
(Sect.~\ref{sec:zcat}).  We choose the peak value of the Probability
Density Function as an estimate of its photo-$z$.  The mean and median
values of the photo-$z$ of the shear catalogue (non-zero weight) are
0.87 and 0.83, respectively. We fit the redshift distribution using
the following formula:
\begin{equation}
\label{eq:photoz}
  p(z)  = A \frac{z^a+z^{ab}}{z^b+c},
\end{equation}
where the best fit values of the parameters $A, a,b,c$ are 0.50, 
0.39, 4.66, 0.60, respectively. The histogram and the fitted
photo-$z$ distributions are shown in Fig.~\ref{fig:photoz}.  
The fitted  
 redshift distribution (Eq.~\ref{eq:photoz}) is used 
 to predict the  shear two-point correlation 
in Sect.~\ref{sect:2pt}. The normalized histogram of photo-$z$ is used
for cosmological constraints (Sect.~\ref{sec:cosmo}) to avoid the
possible bias due to the model fitting.
  
\begin{figure}

  \resizebox{0.9\hsize}{!}{
    \includegraphics[bb = 20 125 580 660 , angle = -90 ]{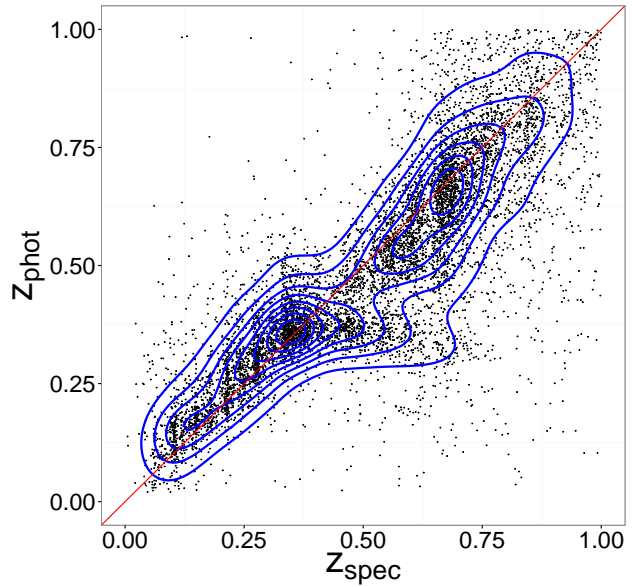}
  }

  \caption{ The photo-$z$ for the galaxies of shear catalogue are
    matched with spectroscopic redshift sample. The contours present
    the density of the galaxy number.  }
   \label{fig:photozshift}
\end{figure}

We note that this paper focuses on presenting the VOICE shear
measurement results.  The photo-$z$ distribution of the background
galaxies are needed for cosmological constraints. We checked the
photo-$z$ measurements by comparing with a subsample with
spectroscopic redshifts (spec-$z$).  We matched the galaxies to the
spectroscopic redshift sample
\citep{2010A&A...518L..20V,2015fers.confE..27V} and found  23638
galaxies.  As shown in Fig.~\ref{fig:photozshift}, the photo-$z$ has
generally a good agreement with spec-$z$. The median value of $\delta z
= ($photo-$z-$spec-$z)/(1+$spec-$z)$ is  $-0.008$
with Median Absolute
Deviation (MAD) value 0.060.  We separated the full sample into two redshift
bins according to the median value 0.83 of the full shear catalogue.
The matched galaxies in low and high bins are  19389 and 4069,
respectively. The sub-samples of two redshift bins show opposite
$\delta z$ as compared to the spectroscopic redshift. We found  $\delta
z = -0.012$ and $0.022$ for the low- and high-$z$ bin. The MAD values are
 0.055 and 0.104, respectively.

 Our photo-z measurements are based on the VOICE $u,g,r,i$ data
  together with four additional near infrared-band data $Y,J,H,K_s$
  (8-band photo-$z$). In the appendix, we compare the photo-$z$ values
  withe the ones determined using only the 4 optical bands (4-band
  photo-$z$), to demonstrate the importance of the near--infrared
  bands.

  \subsection{VOICE-like simulation}
  \label{sec:simulation}

VOICE is about one magnitude deeper than CFHTLenS and KiDS, composed
of a few tens usable exposures for each field.  We need to optimize
$\lensfit$ in order to deal with the high density of background
galaxies and check its capability to work with such a large number of
exposures simultaneously for each galaxy shape measurement.

To validate our optimization and calibrate the measured shear, we
performed image simulations representing the observed $r$-band images.
We briefly summarize the simulation results here and refer to the
paper by \citet{2017LiuDZ} for more details. In the simulation, we use
the sources detected in the stacked images as the input parent sample,
and fix many observing conditions, such as the dithering pattern,
background noise, celestial positions and brightness of the detected
objects, to mimic the real observations.  In this case, galaxy
clustering and blending effect are included naturally.  The $\psfex$
package \citep{2011ASPC..442..435B} was used to model the
spatially-varing PSF for every exposure. For each galaxy, a randomly
sampled intrinsic ellipticity value and a constant shear with modulus
of the reduced shear $|g|$\,=\,0.04 was assigned. In total, four
different shear combinations ($g_1$, $g_2$) were used, namely:
(0.0283, 0.0283), ($-$0.0283, $-$0.0283), (0.0153, $-$0.0370), and
($-$0.0370, 0.0153), respectively. The simulated single exposure
images were then generated by the \textsc{Galsim} toolkit
\citep{2015A&C....10..121R}, and the galaxy shapes were also measured
by $\lensfit$.  Overall, our simulations present good agreements with
the observations, especially the distributions of the PSF
properties. We applied the bin-matching method to the
signal-noise-ratio (SNR) and size plane to calibrate the bias of the
simulation data. The final residual multiplicative bias after
calibration reaches an accuracy of 0.03 with negligible addictive bias
in different SNR and size bins.

The sensitivity of the bias calibration to the undetected and
neighboring objects is also discussed in \citet{2017LiuDZ}. The
undetected objects are likely to skew the background noise so that
they can potentially bias the shape measurements of galaxies,
especially those with low SNR. Taking the depth and noise level into
account, we find that the impact of the undetected galaxies is
negligible for the VOICE survey. Additionally, the bias results from
galaxy blending effect are also analyzed.  Further analyses show that
their impact on the two-point correlation function can be securely
neglected due to the small fraction they account for
(Sect.~\ref{sect:blender}).

\section{Shear two-point correlation analyses}
\label{sect:shear}
Cosmic shear is the weak lensing effect caused by the large-scale
structures in the Universe.  We briefly summarize the theoretical
relations between second-order weak lensing observables and
cosmological quantities in Sect.~\ref{sec:cs_theory}, and then present
the correlation analyses of the VOICE shear catalogue.  For details on
the theoretical foundation of weak gravitational lensing we refer to
the literature \citep[e.g.][]{2001PhR...340..291B,
  2014RAA....14.1061F, 2015RPPh...78h6901K, 2017arXiv171003235M}.

\subsection{Theoretical background}
\label{sec:cs_theory}
Weak lensing induced by the large-scale structures measures the
convergence power spectrum $P_\kappa$ through two-point correlation
statistics.  It is a projection of the total matter density
fluctuation power spectrum $P_\delta$ under the Limber approximation
\citep{1992ApJ...388..272K}:
\begin{equation}
  P_\kappa(\ell) = \int_0^{\chi_{\rm lim}} \dd \chi \,  G^2(\chi) \,
    P_\delta\left(k = \frac{\ell}{f_K(\chi)}; \chi\right).
\end{equation}
The projection integral is carried out over the comoving distances $\chi$, from the
observer out to the limiting distance $\chi_{\rm lim}$ of the
survey. The lensing efficiency $G$ is given by
\begin{equation}
  G(\chi) = \frac 3 2 \left(\frac{H_0} c \right)^2
  \frac{\Omegam}{a(\chi)} \int_\chi^{\chi_{\rm lim}} \dd \chi^\prime p(\chi^\prime)
  \frac{f_K(\chi^\prime - \chi)}{f_K(\chi^\prime)},
\end{equation}
where $H_0$ is the Hubble constant, $c$ is the speed of light,
$\Omegam$ is the present total matter density, and $a(\chi)$ is the
scale factor at comoving distance $\chi$. The cosmology-dependent
comoving angular diameter distance is denoted by $f_K$.

Cosmic shear two-point correlation functions (2PCFs) are the Hankel
transforms of the convergence power spectrum $P_\kappa$, which can be
written as the linear combinations of the E- and B-mode spectra,
$P_{\rm E}$ and $P_{\rm B}$, respectively
\begin{align}
  \xi_\pm(\vartheta)
  & =  {1\over 2\pi} \int_0^\infty {\rm d}\ell\,  \ell \, \left[ P_{\rm
      E}(\ell) \pm P_{\rm B}(\ell) \right] {\rm J}_{0,4}(\ell\vartheta),
  \label{xipmpkappa}
\end{align}
where ${\rm J}_0$ and ${\rm J}_4$ are the first-kind Bessel functions of
order 0 and 4, corresponding to the components $\xi_+$ and $\xi_-$,
respectively.

In real observations, the most direct measurement of weak gravitational
shear signal is derived from galaxy ellipticity measurements.  The
unbiased 2PCFs $\xi_+$ and
$\xi_-$ are estimated by averaging over pairs of galaxies
\citep{2002A&A...396....1S},
\begin{equation}
\hat{\xi}_{\pm}(\vartheta)=\frac{\sum_{ij}{w_iw_j}[\epsilon_t(\bm{\vartheta_i})\epsilon_t(\bm{\vartheta_j})\pm\epsilon_\times(\bm{\vartheta_i})\epsilon_\times(\bm{\vartheta_j})]}{\sum_{ij}{w_iw_j}}.
\label{xipmestim}
\end{equation}
Here, the sum is performed over all galaxy pairs with angular
separation $\vartheta=|\vec \vartheta_i-\vec \vartheta_j|$ within some
bin around $\vartheta$. $\epsilon_t$ and $\epsilon_\times$ are the
tangential and cross-components of the galaxy ellipticity,
respectively, with respect to the line connecting the two
galaxies. $w_i$ is the weight for the $i$-th galaxy, obtained from the
$\lensfit$.

Assuming General Relativity, weak gravitational lensing only
contributes to an E-mode power spectrum and therefore, a non-detection
of the B-mode is a way to check the quality of shear measurement of
the data. The E-/B-mode shear correlations $\xi_{\rm E,B}$, the
aperture-mass dispersion $\langle M_{\rm ap}^2 \rangle$ and the shear
top-hat rms $\langle |\gamma|^2 \rangle$ are the most popularly used
second-order shear correlations.  The decomposed E- and B-mode
estimators in an aperture of radius $\theta$ can be written as
integrals over the filtered correlation functions of $\xi_+$ and
$\xi_-$ \citep{2002ApJ...568...20C,2002A&A...389..729S}, as follows:
\begin{equation}
X_{\rm E, B} (\theta) = \frac 1 2 \sum_i \vt_i \, \Delta \vt_i
\left[ F_+\left( {\vt_i} \right) \xi_+(\vt_i) \pm
       F_-\left( {\vt_i} \right) \xi_-(\vt_i) \right] ,
     \label{X_EB}
\end{equation}
where $\Delta \vartheta_i$ is the bin width varying with $i$. The
estimators $X_{ \rm E}$ and $X_{\rm B}$ are only sensitive to the E-
and B-mode, respectively, with suitable filter functions $F_+$ and
$F_-$. The detail expressions of other two-point correlations are
referred to Table~1 and Appendix~A of \cite{2013MNRAS.430.2200K}.

\subsection{Multiplicative Bias Correction} 

As shown in Eq.~(\ref{xipmestim}), given an unbiased shear
measurement, 2PCFs $\xi_+$ and $\xi_-$ can be estimated, from an
observational point of view, by averaging over pairs of galaxies.
However, data reduction and shear measurement methods can generate
possible biases. Thus a shear calibration \citep{2012MNRAS.427..146H}
is usually applied to describe the relation between the observed shear
and the true signal, which accounts for a potential additive bias
$c_a$ and a multiplicative bias $m_a$ for the $a$-th component of the
galaxy ellipticity ($a=1,2$),
\begin{equation}
\epsilon^{\mathrm{obs}}_a=(1+m_a)\epsilon^{\mathrm{true}}_a+c_a .
\label{e_mc}
\end{equation}
In our analyses, the additive bias is estimated from the observational
shear catalogue, and found to be consistent with zero, on average at
the level of $\sim 8\times 10^{-4}$ and $\sim 3\times 10^{-5}$ for
$\epsilon_1$ and $\epsilon_2$, respectively. However, the
multiplicative biases are non-negligible. We derived them from on our
image simulations \citep{2017LiuDZ}.  In particular, we obtained the
$m$ values in multiple two-dimensional bins of the galaxy SNR and the
size from simulations analysis. We then applied them to the galaxies
in the observed shear catalogue according to their SNR and size. We
found different values for $m_1$ and $m_2$.  We then had to take into
account the multiplicative bias for $\epsilon_1$ and $\epsilon_2$
separately when calculating the shear 2PCFs, which is different from
previous studies, such as CFHTLenS and KiDS. We derived the
corresponding 2PCFs components taking into account different $m$
values as follows.

Considering a pair of galaxies located at $\vec \vartheta_i$ and $\vec
\vartheta_j$, respectively, their tangential and cross components with
respect to the pair separation $\vartheta_i-\vartheta_j$ are given by
\begin{equation}
\epsilon_t=-R\mathrm{e}(\epsilon \mathrm{e}^{-2\mathrm{i}\phi});
~~\epsilon_\times=-I\mathrm{m}(\epsilon \mathrm{e}^{-2\mathrm{i}\phi}) ,
\label{et_ex}
\end{equation}
where $\phi$ is the polar angle $\vec \vartheta_i- \vec \vartheta_j$. 2PCFs
(Eq.~\ref{xipmestim}) can then be  expressed in terms of  a complex ellipticity quantity composed of two
components, $\epsilon=\epsilon_1+\mathrm{i} \epsilon_2$,
\begin{equation}
\hat{\xi}_{+}(\vartheta)=\frac{\sum_{ij}{w_iw_j}[\epsilon_1(\bm{\vartheta_i})\epsilon_1(\bm{\vartheta_j})]}{\sum_{ij}{w_iw_j}}+\frac{\sum_{ij}{w_iw_j}[\epsilon_2(\bm{\vartheta_i})\epsilon_2(\bm{\vartheta_j})]}{\sum_{ij}{w_iw_j}} ,
\label{xipp}
\end{equation}
\begin{equation}
\begin{split}
\hat{\xi}_{-}(\vartheta)&=\frac{\sum_{ij}{w_iw_j}[\epsilon_1(\bm{\vartheta_i})\epsilon_1(\bm{\vartheta_j})\cos(4\phi)]}{\sum_{ij}{w_iw_j}}\\&+\frac{\sum_{ij}{w_iw_j}[-\epsilon_2(\bm{\vartheta_i})\epsilon_2(\bm{\vartheta_j})\cos(4\phi)]}{\sum_{ij}{w_iw_j}}\\
&+\frac{\sum_{ij}{w_iw_j}[\epsilon_1(\bm{\vartheta_i})\epsilon_2(\bm{\vartheta_j})\sin(4\phi)]}{\sum_{ij}{w_iw_j}}\\&+\frac{\sum_{ij}{w_iw_j}[\epsilon_2(\bm{\vartheta_i})\epsilon_1(\bm{\vartheta_j})\sin(4\phi)]}{\sum_{ij}{w_iw_j}} .
\end{split}
\label{ximm}
\end{equation}
Therefore, we need to introduce four calibration factors $1+K_{ab}$
($a=1,2$ and $b=1,2$) here
\begin{equation}
\begin{split}
1+K_{11}&=\frac{\sum_{ij}{w_iw_j}[(1+m_1(\bm{\vartheta_i}))(1+m_1(\bm{\vartheta_j}))]}{\sum_{ij}{w_iw_j}}
;\\
1+K_{22}&=\frac{\sum_{ij}{w_iw_j}[(1+m_2(\bm{\vartheta_i}))(1+m_2(\bm{\vartheta_j}))]}{\sum_{ij}{w_iw_j}}
;\\
1+K_{12}&=\frac{\sum_{ij}{w_iw_j}[(1+m_1(\bm{\vartheta_i}))(1+m_2(\bm{\vartheta_j}))]}{\sum_{ij}{w_iw_j}}
;\\
1+K_{21}&=\frac{\sum_{ij}{w_iw_j}[(1+m_2(\bm{\vartheta_i}))(1+m_1(\bm{\vartheta_j}))]}{\sum_{ij}{w_iw_j}}
,\\
\end{split}
\label{ximm}
\end{equation}
where $1+K_{12}=1+K_{21}$ considering the pair symmetry. The final calibrated 2PCFs are then obtained by
\begin{equation}
\begin{split}
\hat{\xi}_{+}(\vartheta)&=\frac{1}{1+K_{11}}\frac{\sum_{ij}{w_iw_j}[\epsilon_1^{\mathrm{obs}}(\bm{\vartheta_i})\epsilon_1^{\mathrm{obs}}(\bm{\vartheta_j})]}{\sum_{ij}{w_iw_j}}\\&+\frac{1}{1+K_{22}}\frac{\sum_{ij}{w_iw_j}[\epsilon_2^{\mathrm{obs}}(\bm{\vartheta_i})\epsilon_2^{\mathrm{obs}}(\bm{\vartheta_j})]}{\sum_{ij}{w_iw_j}};
\end{split}
\label{xipp_final}
\end{equation}
\begin{equation}
\begin{split}
\hat{\xi}_{-}(\vartheta)&=\frac{1}{1+K_{11}}\frac{\sum_{ij}{w_iw_j}[\epsilon_1^{\mathrm{obs}}(\bm{\vartheta_i})\epsilon_1^{\mathrm{obs}}(\bm{\vartheta_j})\cos(4\phi)]}{\sum_{ij}{w_iw_j}}\\&+\frac{1}{1+K_{22}}\frac{\sum_{ij}{w_iw_j}[-\epsilon_2^{\mathrm{obs}}(\bm{\vartheta_i})\epsilon_2^{\mathrm{obs}}(\bm{\vartheta_j})\cos(4\phi)]}{\sum_{ij}{w_iw_j}}\\
&+\frac{1}{1+K_{12}}\frac{\sum_{ij}{w_iw_j}[\epsilon_1^{\mathrm{obs}}(\bm{\vartheta_i})\epsilon_2^{\mathrm{obs}}(\bm{\vartheta_j})\sin(4\phi)]}{\sum_{ij}{w_iw_j}}\\&+\frac{1}{1+K_{21}}\frac{\sum_{ij}{w_iw_j}[\epsilon_2^{\mathrm{obs}}(\bm{\vartheta_i})\epsilon_1^{\mathrm{obs}}(\bm{\vartheta_j})\sin(4\phi)]}{\sum_{ij}{w_iw_j}} .
\end{split}
\label{ximm}
\end{equation}
%

\subsection{Shear two-point correlation estimations}
\label{sect:2pt}

\begin{figure*}

  \resizebox{0.9\hsize}{!}{
    \includegraphics[bb = 30 20 570 400]{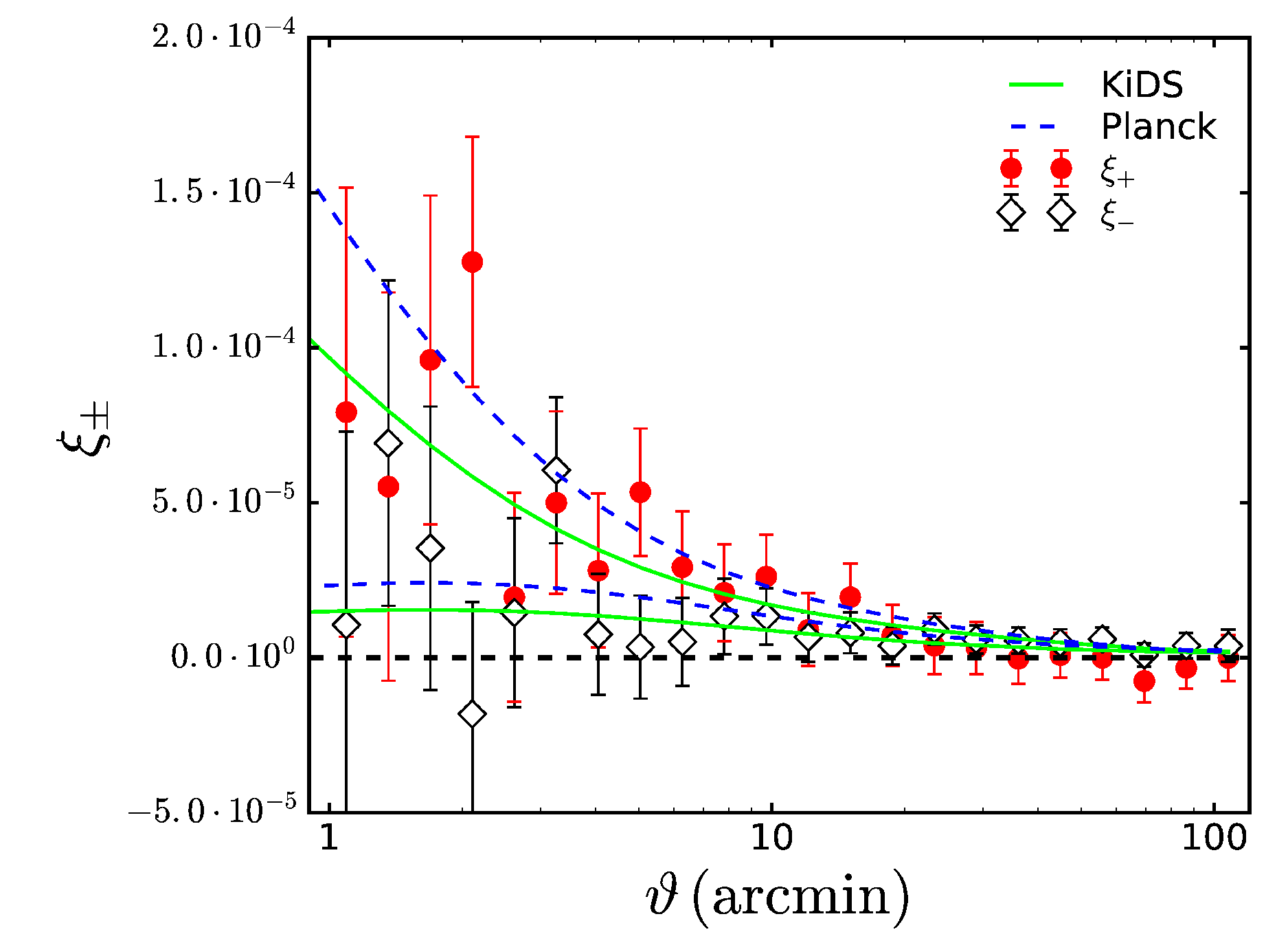}
    \includegraphics[bb = 30 20 570 400]{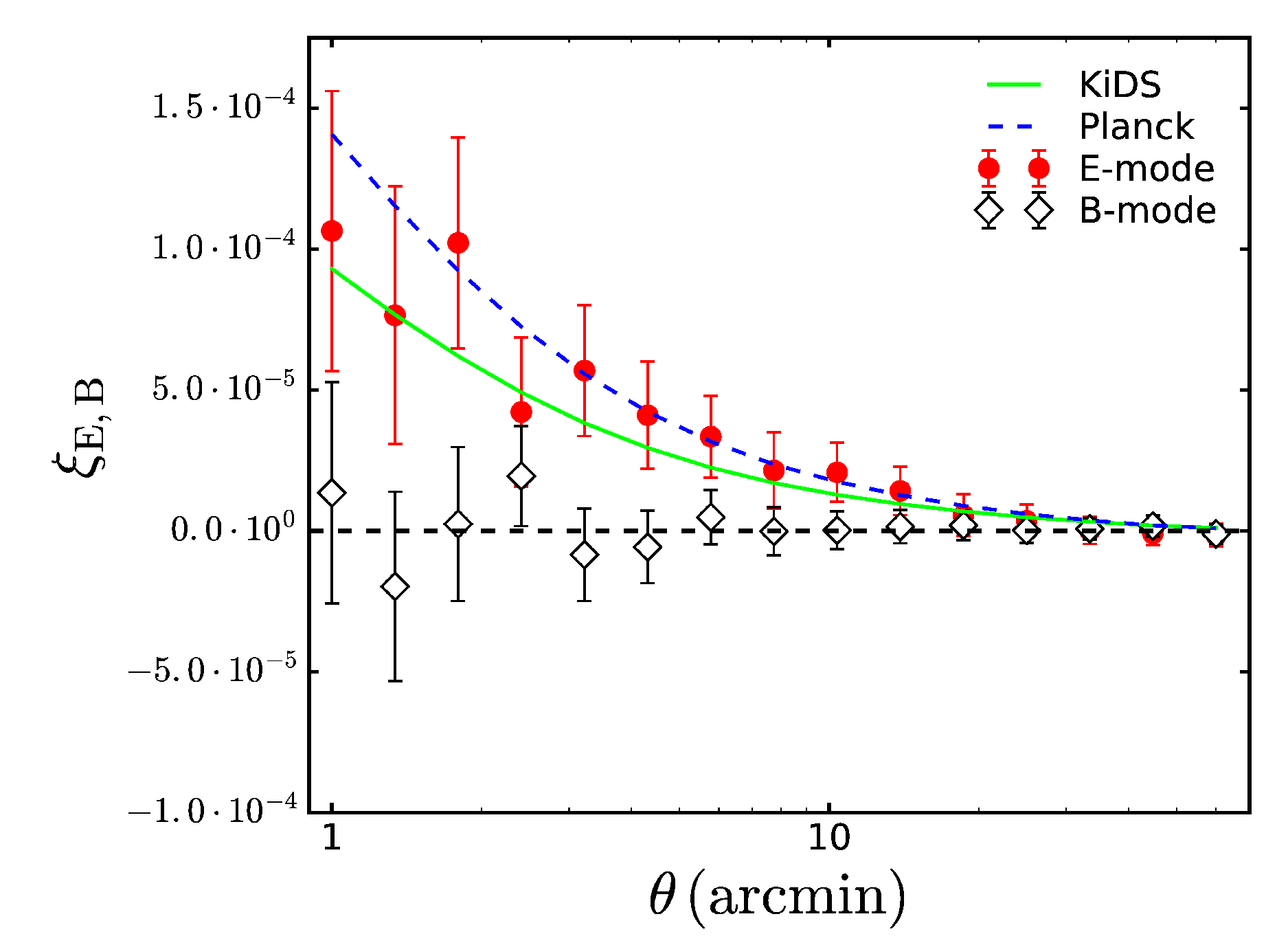}
  }
 \resizebox{0.9\hsize}{!}{
    \includegraphics[bb = 30 20 570 420]{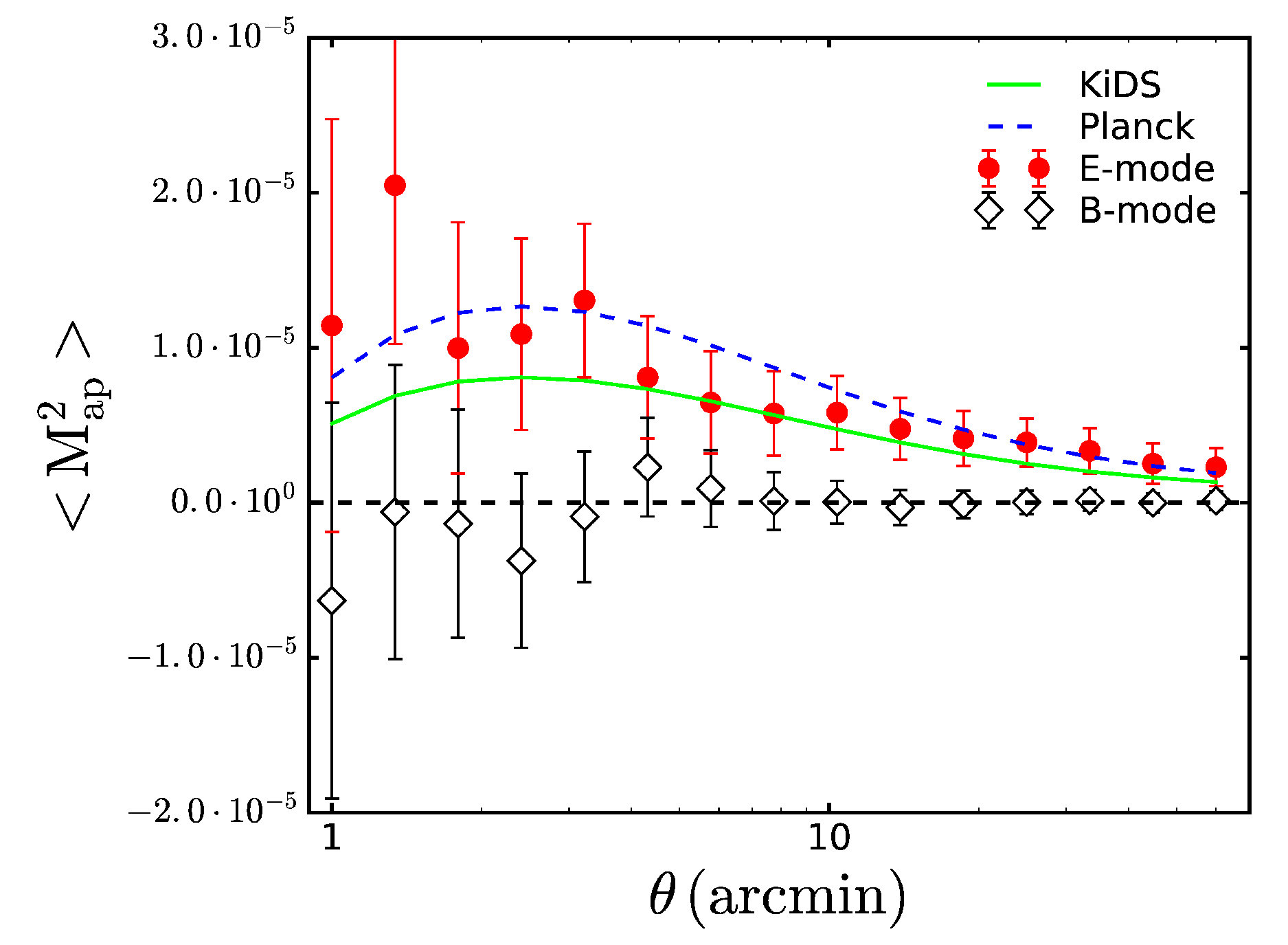}
    \includegraphics[bb = 30 20 570 420]{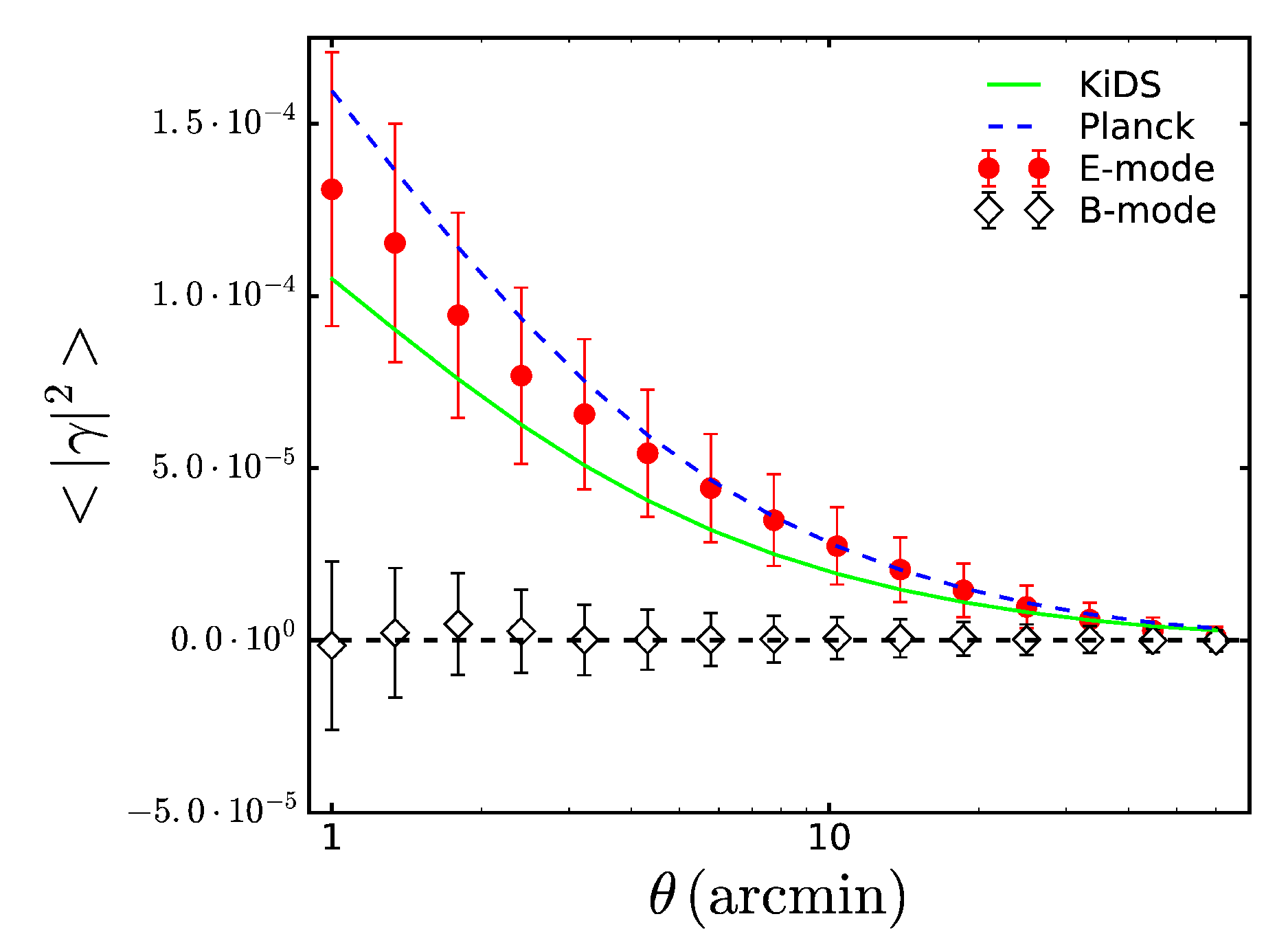}
  }
  
  \caption{ The calibrated shear correlation functions of the combined
    four tiles of VOICE-CDFS: {\it top-left panel}: $\xi_+$ (red full
    dots) and $\xi_-$ (black open diamonds). The angular distance
    $\vartheta$ is the seperation between the galaxy pairs; {\it
      top-right panel}: $\xi_{\rm E,B}$; {\it bottom-left panel}:
    $\langle M_{\rm ap}^2 \rangle$; {\it bottom-right panel}: $\langle
    |\gamma|^2 \rangle$. They are the derived 2PCFs with an aperture
    of radius $\theta$, where E-modes are full dots and B-modes are
    black open diamonds.  The error bars correspond to square root of
    the covariance diagonal term. Two theoretical predictions based on
    the cosmological model from KiDS (green solid line) and Planck15
    (blue dash line) are shown using the VOICE photo-$z$ distributions
    (see Eq.~\ref{eq:photoz}).  }
   \label{fig:2pt}
\end{figure*}

Based on the above analyses, we computed the shear 2PCFs using the
combined VOICE shear catalogue from the four CDFS tiles. The results
are shown in Fig.~\ref{fig:2pt}. The upper left panel shows $\xi_{+}$
(red full dots) and $\xi_{-}$ (black open diamonds), respectively. The
upper limit of the angular separation considered here is taken to be
$120\arcmin$, as the survey area is $2\times 2\hbox{ deg}^2$. For the
lower limit, although we show the results from $\vartheta=1\arcmin$ in
Fig.~\ref{fig:2pt}, we actually calculate $\xi_\pm$ starting from
$10\arcsec$, which corresponds to the $\lensfit$ postage stamp size
(48 pixels).

The other three panels in Fig.~\ref{fig:2pt} show the results of
$\xi_{\rm E,B}$ (top-right), $\langle M_{{\rm ap}}^2 \rangle$
(bottom-left) and $\langle |\gamma|^2 \rangle$ (bottom-right),
respectively. They are derived from $\xi_\pm$ by performing
integrations with different filters. To avoid introducing artificial
B-mode due to the finite integration range, we considered these three
quantities only up to the angular scale $\theta=60\arcmin$, the radius
of an aperture with maximum separation in a galaxy pair.  It is seen
that the B-mode is consistent with zero for all the three derived
quantities in the given angular range.  The multiplicative biases of
$\xi_\pm$ have been corrected (Eq.~\ref{xipp_final} and \ref{ximm}).
The amplitudes of the corrections on 2PCFs are in the order of a few
percent.

The different filter functions of three derived second-order functions
lead to different sensitivities on smoothing scales. For instance,
$\langle |\gamma|^2 \rangle$ is the one with the highest correlation
between data points, thus the E-/B-mode components look smoother than
those of the other two quantities.  The error bars are the squared
root of the diagonal terms of the covariance matrix measured from
VOICE-like ray-tracing simulations to be described in
Sect.~\ref{sect:cv}.

The results are compared to the theoretical predictions using the
cosmological parameters derived from KiDS \citep{2017MNRAS.465.1454H}
and Planck15 \citep{2016A&A...594A..13P}, where
$\Omegam=0.231 ;\sigma_8=0.851$ and $\Omegam=0.315 ;\sigma_8=0.831$
respectively, with the same angular scale range $[10\arcsec,
  120\arcmin]$ for $\xi_\pm$. 
The redshift distribution used for the
theoretical predictions 
 is obtained by fitting the Eq.~(\ref{eq:photoz}) 
to the photo-$z$ distribution of the VOICE shear catalogue,
and is shown as a solid line in
Fig.~\ref{fig:photoz}.

\subsection{Covariance Estimation}
\label{sect:cv}

To model and interpret the observed 2PCFs, we need to estimate the
error covariance. To do so, we used the N-body simulations described
in \cite{2015MNRAS.450.2888L} to account for the non-Gaussianity of
the cosmic shear field on small and medium angular scales, and
performed ray-tracing calculations to construct the shear and
convergence maps. The cosmology involved is the flat $\Lambda$ cold
dark matter ($\Lambda$CDM) model with $\Omega_\mathrm{m}=0.28$,
$\Omega_\Lambda=0.72$, $\Omega_\mathrm{b}=0.046$, $\sigma_8=0.82$,
$n_s=0.96$ and $h=0.7$, where $\Omega_{\rm m}$, $\Omega_{\Lambda}$,
and $\Omega_{\rm b}$ are the present dimensionless densities of the
total matter, cosmological constant, and the baryonic matter,
respectively, $\sigma_8$ is the rms of linearly extrapolated density
perturbations over $8\hbox{ Mpc}h^{-1}$, $n_s$ is the power index of
the power spectrum of initial density fluctuations, and $h$ is the
Hubble constant in units of $100\hbox{ km/s/Mpc}$.  In order to cover
a large redshift range up to $z=3$ in ray-tracing calculations, we
padded 12 independent simulation boxes, with 8 small boxes each with a
size of $320\hbox{ Mpc}h^{-1}$ to $z=1$ and 4 larger boxes each with a
size of $600\hbox{ Mpc}h^{-1}$ from $z=1$ to $z=3$, and used in total
59 lens planes. From one set of padded boxes, we can generate 4 sets
of lensing maps each with an area of $3.5\times 3.5 \deg^2$ sampled on
$1024\times 1024$ pixels. For each set, we have 59 shear and 59
convergence maps at 59 different redshifts corresponding to the far
edges of the 59 lens planes. In total, we run 24 sets of simulations,
and generate lensing maps with the total area of $1176\deg^2$.  A more
detailed descriptions for our N-body simulations and ray-tracing
calculations can be found in \cite{2015MNRAS.450.2888L} and
\cite{2014ApJ...784...31L}.

With these lensing maps, we then generated 384 VOICE-like mock catalogues
to estimate the error covariance. The generating procedure for each
mock is as follows.

(i) We placed the 4 continuous VOICE tiles randomly over the simulated
map area, with the positions, photo-$z$, galaxy weights
and the mask information preserved in the analyses. The amplitudes of
ellipticities of the galaxies  were also preserved, but with their
orientations being randomized.

(ii) For each galaxy in the catalogue, its reduced shear $\bm{g}$  was
calculated by interpolating the signals from the pixel positions on
simulated maps to the galaxy position. The interpolation  was also done
in  redshift. Regarding the randomized ellipticity
obtained in (i) as its intrinsic ellipticity $\bm{\epsilon_s}$, the
mock observed ellipticity $\bm{\epsilon}$ can then be constructed from
\begin{equation}
\boldsymbol {\epsilon}(\boldsymbol \vartheta, z)=\left\{ \begin{array}{ll} \frac{\boldsymbol{\epsilon}_s(\boldsymbol \vartheta, z)+
\boldsymbol{g}(\boldsymbol \vartheta, z)}{1+\boldsymbol{g^*}(\boldsymbol \vartheta, z)\boldsymbol{\epsilon}_s(\boldsymbol \vartheta, z)}
& \textrm{for $\vert{\boldsymbol{g}(\boldsymbol \vartheta,z)}\vert\leq 1$}\\ \\ \frac{1+\boldsymbol{g}(\boldsymbol \vartheta, z)
\boldsymbol{\epsilon}_s^{*}(\boldsymbol \vartheta, z)}{\boldsymbol{\epsilon}_s^{*}(\boldsymbol \vartheta, z)+\boldsymbol{g^*}(\boldsymbol \vartheta, z)}
& \textrm{for $\vert{\boldsymbol{g}(\boldsymbol \vartheta,
      z)}\vert>1$}  
 \end{array}\right ..
\label{eobs}
\end{equation}

(iii) The 2PCFs analyses were then carried out for each mock, with the
same procedures for the observed data, the error covariance can be
further estimated with these 2PCFs results from the whole 384 mocks.
These covariance matrices  were used to give error bars shown in Fig.9,
and also applied to derive cosmological constraints to be presented in
Sect.~\ref{sec:cosmo}.

\subsection{The star-galaxy cross-correlation function}
\label{sect:sgcor}

  The results in Fig.~\ref{fig:2pt} show that our VOICE shear catalogue
exhibits no detectable B-mode. To further check the data quality, we
analyze the level of PSF-related systematics by measuring the
star-galaxy cross correlation  $\bm{\xi}_{\rm sg}(\vartheta) = \langle
\epsilon^{\rm obs} \bm{e^\star} \rangle$,
where
$\epsilon^{\rm {obs}}$ is the observed shear estimators, $\bm
{e^\star}$ 
is
a complex $N$ dimensional vector of PSF ellipticity at the position
of the galaxy in each of the $N$ dithered exposures of the field.
For these analyses, star-galaxy pairs with the angular separation
$\vartheta$ in the range of $[1\arcmin, 60\arcmin]$   were taken into account,
and they  were divided into 6 evenly-distributed log-normal bins. The
zero-lag star-galaxy correlation $\bm{\xi}_{\rm sg}(\vartheta=0)$,
hereafter $\bm{\xi}_{\rm sg}(0)$, which indicates the primary
systematics,  was derived using the model of PSF ellipticity to
determine $\bm{e^\star}$ at the location of each galaxy, with
\begin{equation}
\bm{\xi}_{\rm sg}(0)=\frac{\Sigma{w_i[\epsilon_1(\bm{\vartheta_i})\bm{e_1^\star}(\bm{\vartheta_i})+\epsilon_2(\bm{\vartheta_i})\bm{e_2^\star}(\bm{\vartheta_i})]}}{\Sigma{w_i}}.
\label{xisg0}
\end{equation}
If the PSF model and correction are correct  so that  the observed shear
estimator  is  uncorrelated with  the PSF, $\bm{\xi}_{\rm sg}(0)$ should be
consistent with zero.

Following some arguments discussed in \cite{2012MNRAS.427..146H}, with
a measure of the zero-lag star-galaxy correlation $\bm{\xi}_{\rm
  sg}(0)$, we can make a prediction of the star-galaxy correlation at
any angular scale using
\begin{equation}
\bm{\xi}_{\rm sg}(\vartheta)\approx\bm{C}_0^{-1}\bm{\xi}_{\rm sg}(0)\bm{C}_\vartheta,
\label{xisg_approx}
\end{equation}
where $\bm{C}_0$ is the measured covariance matrix of PSF
ellipticities between exposures at zero-lag and $\bm{C}_\vartheta$ is the
same PSF measurement but for sources at separation $\vartheta$. Here we
only consider the case using weighted PSF ellipticities in the final
shear catalogues. Thus, Eq.~(\ref{xisg_approx}) reduces to
\begin{equation}
\xi_{\rm sg}(\vartheta)\approx\xi_{\rm sg}(0)\langle e_a^\star e_b^\star\rangle/\langle e^{\star2}\rangle,
\label{xisg_approx2}
\end{equation}
where $a$ and $b$ indicate objects separated by a distance $\vartheta$.

\begin{figure}
   \resizebox{0.9\hsize}{!}{
          \includegraphics[bb = -35 140 590 600 ]{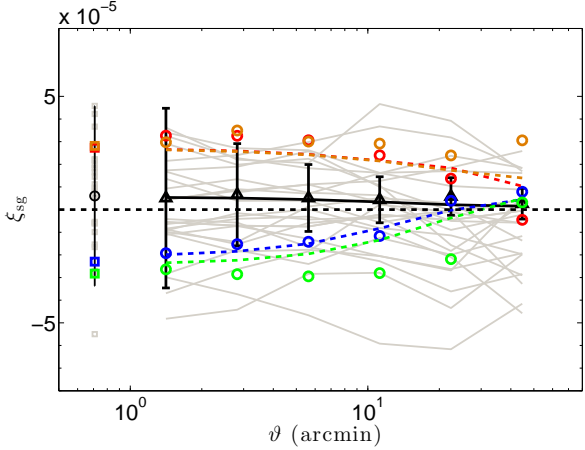}
     }
\caption{The star-galaxy cross-correlation function $\xi_{\rm
    sg}(\vartheta)$ measured in CDFS1-4  (e.g.,  black triangles with
  error bars),  compared to the predicted angular
  star-galaxy correlation (e.g., Eq.~\ref{xisg_approx2},  black solid line)
  calculated using only the zero separation measure $\xi_{\rm sg}(0)$
  (shown offset, gray circle with error bar). The corresponding error
  bars are assigned using the standard deviation of $\xi_+$ at the
  corresponding $\vartheta$ evaluated from the constructed 384
  mocks. Blue, red, green and orange circles without error bars
  are the measured star–galaxy cross-correlation function for CDFS1,
  CDFS2, CDFS3 and CDFS4, respectively. The corresponding squares and
  dash lines are the corresponding zero-lag and predicted measures for
  different individual fields. As a comparison, the bright  grey
  lines are the measured star-galaxy cross-correlation function for
  the 24 G15 fields in  the  KiDS survey. }
\label{fig:xisg}
\end{figure}

Fig.~\ref{fig:xisg} shows the star-galaxy cross-correlation function
$\xi_{\rm sg}(\vartheta)$ measured in CDFS1-4 fields. Generally
speaking, the whole star-galaxy cross-correlation function is
consistent with zero and is well within the range  of values observed in 
the KiDS survey.

\subsection{Tomography check}
\label{sect:tomo}

\begin{figure*}

  \resizebox{0.9\hsize}{!}{
    \includegraphics[bb = 60 60 550 750, angle=-90]{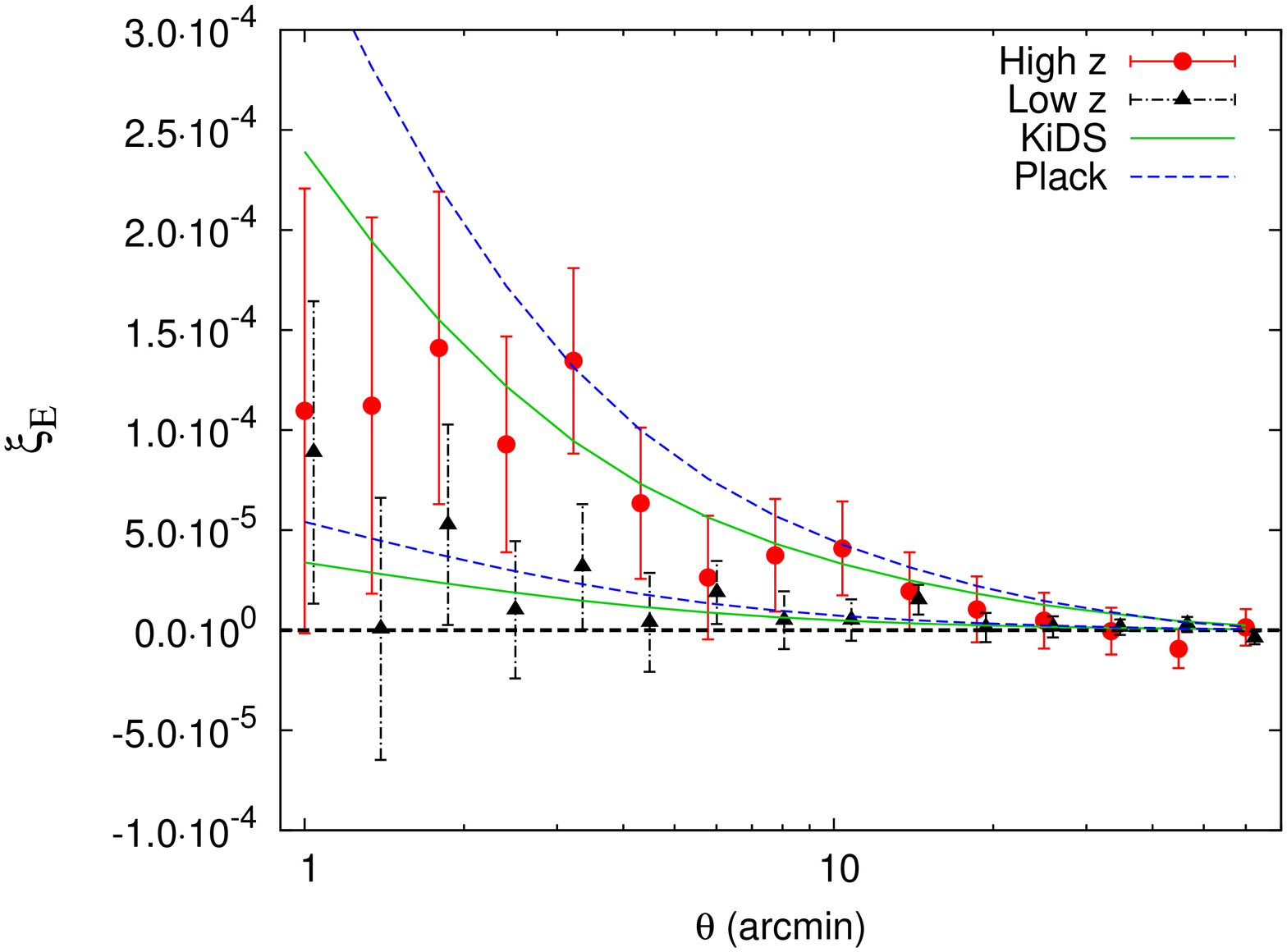}
    \includegraphics[bb = 60 60 550 750, angle=-90]{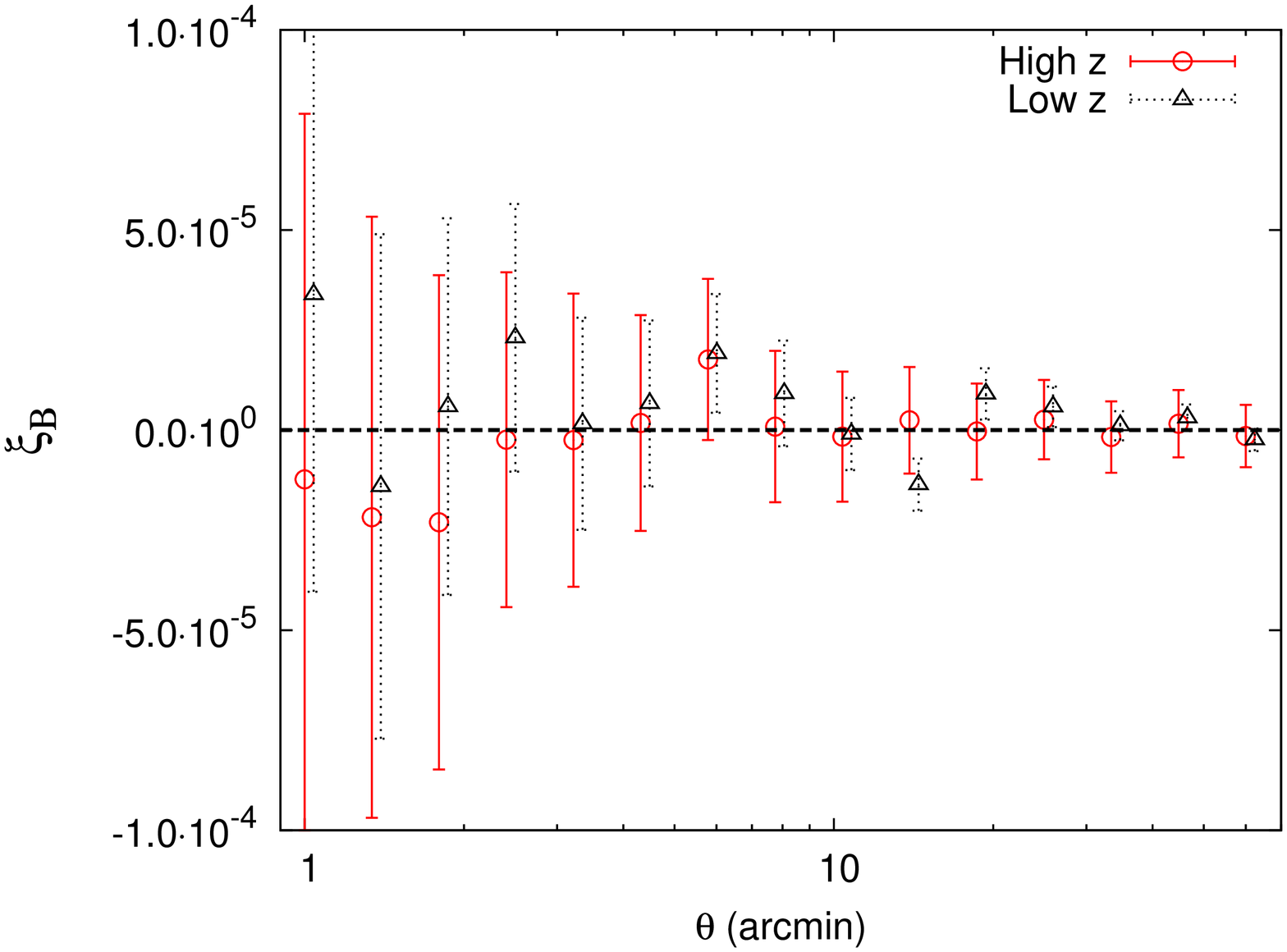}
  }
  \caption{ The calibrated shear correlation functions $\xi_{\rm E}$
    ({\it left panel}) and $\xi_{\rm B}$ ({\it right panel}) of two
    photo$-z$ bin samples. The calculation of error bars and the
    theoretical predictions are the same as those of
    Fig.~\ref{fig:2pt}. The theoretical predictions are estimated
    using the cosmological parameters derived from KiDS (green solid
    lines) and Planck15 (blue dash lines). }
   \label{fig:2pttomo}
\end{figure*}

The reliability of shear measurement in the VOICE data can be further
tested by considering the tomographic shear signals.  We separate the
full shear sample into two photo-$z$ bins divided by the median
photo-$z$ of 0.83.  The results of $\xi_{\rm E}$ (left) and $\xi_{\rm B}$ (right)
are shown in Fig.~\ref{fig:2pttomo}. As expected, the shear
correlation of the high redshift bin is higher than that of the low
redshift bin.  There are no obvious B-modes in all angular scales in
both of cases.  The solid green lines are the theoretical predictions
assuming the KiDS and Planck15 cosmology with the redshift
distributions for the two bins directly from the photo-$z$
measurements.  We can see that our results are in good agreements with
the theoretical predictions.

As this paper mainly focuses on the shear measurement of VOICE, the
tomographic results presented  here are only  for checking the reliability of
the shape measurement. Being our next task, we will perform
cosmological studies using the tomographic correlations from
VOICE. For that, we will consider carefully the impacts of galaxy
intrinsic alignments and photo-$z$ errors.

%

\subsection{Blending Effect}
\label{sect:blender}

The final mosaic reaches  a  $5\sigma$ limiting
magnitude of $r_\mathrm{AB}$\,$\sim$\,26.1\,mag with $2\arcsec$
aperture diameter for point sources. Over 488,000 galaxies are
detected with a number density of 32.85 arcmin$^{-2}$ 
 after excluding the masked regions. 
Following \citet{2013MNRAS.434.2121C}, we define the neighbors simply
by their separation on the celestial sphere. We find that only 0.04\%
of galaxies have neighbors within $1.0\arcsec$, while the fraction
increases dramatically to over 16\% within a 3.0$\arcsec$
separation. These galaxies can be either physically related neighbors
which have similar shear or projected close pairs, with different
redshifts and shape distortions. Though $\lensfit$ has encoded an
algorithm to deal with them \citep{2013MNRAS.429.2858M}, potential
bias is still inevitable in the measured shear due to the
inappropriate modeling of the surface brightness distributions in the
overlapping regions.

\begin{figure}

  \resizebox{0.9\hsize}{!}{
    \includegraphics[bb = 60 60 550 750, angle=-90]{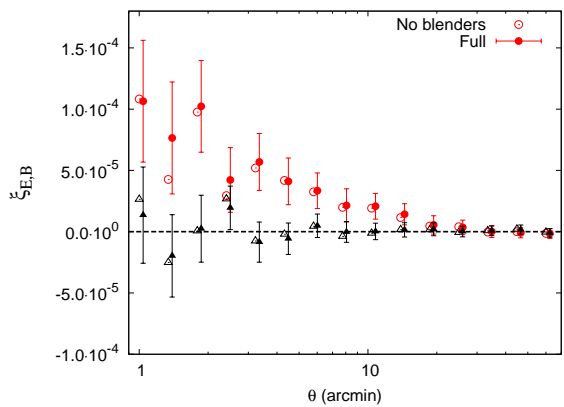}
  }
  \caption{ The calibrated shear correlation function $\xi_{\rm {E,B}}$
    after excluding the blended galaxies (open symbol) is compared to
    that of the full galaxy sample (solid symbol). The E-modes are
    circle in
    red and the B-modes are triangle in black.  The uncertainties are 
    calculated as in      Fig.~\ref{fig:2pt}.  }
   \label{fig:2ptblend}
\end{figure}

Although most of the neighbors have been excluded by $\lensfit$, about
31.6\% of the neighboring galaxies within separation $r=3.0\arcsec$
still have shape measurements.  The ellipticity dispersion of these
remainders is 3.4\% larger than the overall dispersion.   Their weighted
number density  is about 1.28 arcmin$^{-2}$. We compare the
shear two-point correlation functions of the full sample and that  derived
after rejecting  neighbors within $r\leq3.0\arcsec$. The results are shown in
Fig.~\ref{fig:2ptblend}. We find that the differences are within the
error bars given the relatively large statistical uncertainties of the
VOICE shear sample. For future large surveys with dramatically reduced
statistical errors, the neighboring contaminations need to be
carefully accounted for.

From our image simulations \citep{2017LiuDZ}, we further quantify the
impact of the close neighbors on the multiplicative biases. It is
found that the SNR of these galaxies are systematically overestimated
by $\lensfit$ due to the contamination of the neighboring galaxy. As a
result, these close neighbors do provide an additional contribute to
the multiplicative bias, especially at high SNR. The weighted average
bias resulting from these neighbors is about 0.002 from our simulation
analyses. Although this can be safely neglected for the VOICE
analyses, it can be a serious concern for future large surveys that
need the multiplicative bias to be controlled at the level less than
0.001.

%

%
  \section{Cosmological Constraints}
  \label{sec:cosmo}

The most sensitive constraints from weak lensing alone are the
cosmological parameters of the matter density $\Omegam$ and the linear
amplitude of mass fluctuations $\sigma_8$.  In this section, we
present the marginalized constraints for $\Omegam$ and $\sigma_8$  in a
flat $\Lambda$CDM cosmological model.  We note that the main focus of
the paper is to present the shear measurements. The cosmological
constraints shown here are  presented as a reliability check,  in addition to
the 2PCFs presented in the previous sections. Considering also the
relatively large statistical uncertainties of the VOICE shear catalog,
here we do not discuss different possible systematics, such as galaxy
intrinsic alignments, baryonic effects, photo-$z$ errors, etc.. We
will do more careful cosmological analyses as our next task.

\subsection{Sampling the posterior}

We use the open source code
\texttt{Cosmo\_PMC}\footnote{http://cosmopmc.info}
\citep{2011arXiv1101.0950K} to sample the VOICE weak-lensing
constraint posterior with Population Monte Carlo (PMC).  For the flat
$\Lambda$CDM model, the base parameters are $\Omegam$, $\sigma_8$,
$\Omega_{\rm b}$, $n_{\rm s}$ and $h$.  The prior ranges are
summarized in Table~\ref{tab:pars}.

The perplexity parameter $p$ of \texttt{Cosmo\_PMC} is a value between
0 and 1, where 1  stay for a perfect agreement between importance
function and the posterior. Generally, $p$ reaches 0.7 after 10  iterations, 
 after which we stopped the iterations. 
We used 30,000 sample points in each iteration. For the last iteration, larger
samples with 300,000 points are used to reduce the Monte-Carlo
variance.

 \begin{table}
 \begin{center}
 \caption{The parameters sampled under the weak-lensing posterior.
   The second column indicates the (flat) prior ranges analyzed with flat $\Lambda$CDM.}
 
\begin{tabular}{@{}lll}
\hline
Param.\           & Prior & Description  \\ \hline
$\Omegam$       & $[0; 1.2]$  & Total matter density \\
$\sigma_8$      & $[0.2; 1.5]$ & Power-spectrum normalisation\\
$\Omega_{\rm b}$ & $[0; 0.1]$  & Baryon density \\
$n_{\rm s}$      & $[0.7; 1.2]$ & Spectral index of prim.\ density fluct.\  \\
$h$             & $[0.4; 1.2]$ & Hubble parameter\\  
\hline \\
 \end{tabular}
 \label{tab:pars}
\end{center}

 \end{table}

\subsection{Choice of  second-order estimators}

We mainly use the aperture mass dispersion $\langle M_{\rm
  ap}^2\rangle$ for deriving cosmological constraints, for the
following reasons. 1) The filter function of $\langle M_{\rm
  ap}^2\rangle$ is much narrower compared to the one of top-hat shear
rms $\langle |\gamma|^2 \rangle$. Thus $\langle M_{\rm ap}^2\rangle$
of different smoothing scales $\theta$ are less correlated. 2) For
$\langle M_{\rm ap}^2\rangle$, only the lower angular limit is
problematic and causes leakage of the B-mode into the E-mode signal on
small smoothing scales.

\cite{Anderson2003} and \cite{2007A&A...464..399H} have shown that the
inverse covariance calculated directly from the covariance matrix
constructed from simulations is biased, resulting  in  a biased maximum
likelihood (ML) estimator.  We correct the ML estimator by multiplying
 per the Anderson-Hartlap factor $A = (n-p-2)/(n-1)$
\citep{2007A&A...464..399H}.  The bias depends on the number of
simulations $n$ and the number of data bins $p$. Here we have $n=384$
and $p=15$.  Thus the correct factor is $A = 0.96$.

\begin{figure}
  \resizebox{0.9\hsize}{!}{
    \includegraphics[bb = 10 10 350 325]{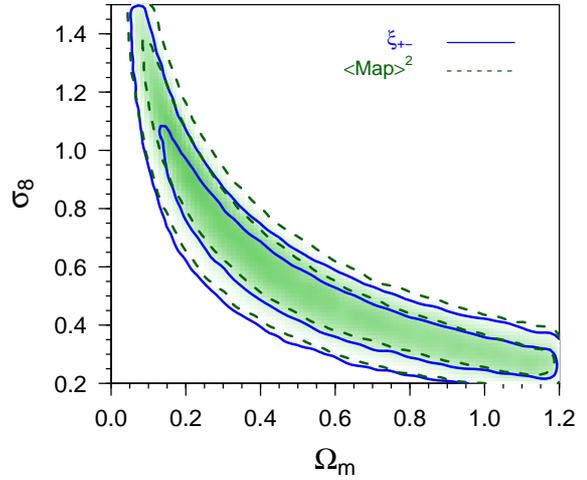}
  }
  \caption{Marginalized posterior density contours (68.3 per cent and  95.5 per
cent) for $\Omegam$ and $\sigma_8$ are constrained from 
    $\xi_{\pm}$ and $\langle M_{\rm ap}^2\rangle$ in the case
    of flat $\Lambda$CDM.}
   \label{fig:Lencom}
\end{figure}

Before presenting the main constraints, we first check the consistency
by comparing the constraints from $\langle M_{\rm ap}^2\rangle$ and
those from the 2PCFs $\xi_{\pm}$ for the flat $\Lambda$CDM model. The
results are shown in Fig.~\ref{fig:Lencom}. It is seen that the two
second-order quantities give rise to very similar contours in the
plane of $\Omegam$ and $\sigma_8$.  This demonstrates that the B-mode of
$\langle M_{\rm ap}^2\rangle$ has negligible impact on the
cosmological parameters constraints.

\subsection{Results}
\label{sect:results}
\begin{figure}

  \resizebox{0.9\hsize}{!}{
    \includegraphics[bb = 10 10 350 325]{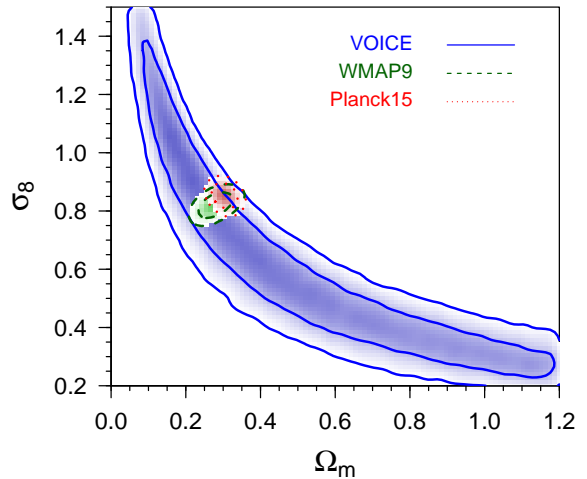}
  }
  
  \caption{ Marginalized posterior density contours (68.3 per cent,
    95.5 per cent) of $\Omegam$ and $\sigma_8$ for  flat
    $\Lambda$CDM from VOICE weak lensing (blue), WMAP9 (green) and
    Planck15 (red).  }
   \label{fig:sig8-Omgm}
\end{figure}

The goal of this paper is to present the VOICE shear catalog
measurements, which we have used to obtain the marginalized
constraints of $\Omegam$ and $\sigma_8$ for flat $\Lambda$CDM
cosmological model in Fig.~\ref{fig:sig8-Omgm}. The degeneracy
direction of these two parameters is approximately a power law, while
its amplitude is given by the parameter $\Sigma_8 =
\sigma_8(\Omegam/0.3)^{\alpha}$.

In order to compare to the results from the KiDS analyses, we fix
$\Omegam = 0.3$ and derive the constraints of $\Sigma_8$ and
$\alpha$. We obtain $\Sigma_8 = 0.70^{+0.11}_{-0.12}$ and $\alpha =
0.64\pm0.02$ assuming a $\Lambda$CDM model, while by fixing
$\alpha=0.5$, as done for KiDS-450 \citep{2017MNRAS.465.1454H}, we
obtain $\Sigma_8=0.68^{+0.11}_{-0.15}$. These results are in broad
agreements with the ones from KiDS-450 and from other literature,
showing that our shear measurements are not affected by systematics
comparing to the statistical uncertainties.

Finally, we compare these results with constraints derived from CMB
measurements from
WMAP9\footnote{https://lambda.gsfc.nasa.gov/product/map/dr5/parameters.cfm}
(green) and
Planck15\footnote{https://wiki.cosmos.esa.int/planckpla2015} (TT +
lowP, red) in Fig.~\ref{fig:sig8-Omgm}. The VOICE constraints are in
broad agreements with both, due to the relatively large statistical
uncertainties.  However, we note that, despite being statistically
consistent, a mild offset with PLANCK15 can still be seen, which goes
in the same direction of the tension found by KiDS-450. A similar
tension is seen if we compare with Planck polarization data (TT $+$ TE
$+$ EE $+$ lowP), again despite the large statistic error of VOICE
shear 2PCF.

To conclude, the above analyses mainly show the validity of our shear catalog
and  the consistency with other results based on wider but shallower datasets.
The detailed cosmological studies taking into account different
systematics will be presented in  a forthcoming paper.

%

\section{Summary}
\label{sec:summary}
  We have presented the cosmic shear measurement of the 4.9 deg$^2$
  CDFS field from $r$-band images of the VOICE survey at the
  VST/OmegaCAM.  Each of the four pointings covering the area has been
  observed with more than 100 exposures. After a stringent selection
  for high quality data, including cuts on seeing and sky background
  brightness variation, about one-third of the exposures have been
  used to obtain the shear measurement.  The final $r$-band co-added
  image reaches a $r = 26.1$ $5\sigma$ limiting magnitude for point
  sources, which is 1.2 mag deeper than KiDS.  We have used the
  software $\lensfit$ to measure the galaxy shapes, which was
  successfully applied on CFHTLenS and KiDS. The novelty of our
  approach, though, is that this is the first time that $\lensfit$ is
  applied to a deep survey with more than a few tens exposures. To
  check the accuracy of our shear measurement we have used VOICE-like
  imaging simulations, which have been fully illustrated in a
  companion paper \citep{2017LiuDZ}. From the mock observations, we
  have obtained the multiplicative bias calibration values at
  different galaxy SNR and size bins to correct the real measurements.
  After these calibrations, the final residual multiplicative bias of
  $\lensfit$ shear measurement is measured with an accuracy of 0.03
  with negligible addictive bias. The final VOICE-CDFS shear catalogue
  contains more than $3\times 10^5$ galaxies with non-zero weight,
  corresponding to the effective number density of galaxies of 16.35
  arcmin$^{-2}$, about twice the one of KiDS.  The photo-$z$ of each
  galaxy have been estimated using the VOICE $u,g,r,i$ together with
  the near-infrared $Y,J,H,K_s$ VIDEO data.  The mean redshift of the
  shear catalogue is 0.87, considering shear weights.

  To check the reliability of the VOICE shear catalogue, we have
  calculated the star-galaxy cross-correlations.  Generally speaking,
  the whole star-galaxy cross-correlation function has been found
  consistent with zero.  We further calculated the 2D shear 2PCFs and
  the derived second-order statistics, and those with two tomographic
  redshift bins divided by the median redshift 0.83 of the sample. The
  results are in agreement with the theoretical predictions using the
  cosmological parameters derived from KiDS and Planck15.
  
  VOICE is a deep imaging survey, and it is important to assess the
  impact of possible blending effect. As discussed in detail in
  \cite{2017LiuDZ}, although most of the neighbours have been excluded
  by $\lensfit$, about 31.6\% of the neighbouring galaxies within
  separation $r=3.0\arcsec$ still have shape measurements. By
  comparing the shear two-point correlation functions between the full
  sample and that after rejecting $r\leq3.0\arcsec$ neighbors, we have
  found that the impact of these neighbouring galaxies on the shear
  correlations is within the VOICE statistical uncertainties.  This
  can be a serious concern, however, for future large and deep
  surveys.

To further validate our shear measurements, we have derived
cosmological constraints from the second-order shear statistics
$\langle M_{\rm ap}^2\rangle$.  We have shown the marginalized
constraints for $\Omegam$ and $\sigma_8$ of flat $\Lambda$CDM
cosmological model, which has found to be $\Sigma_8 =
\sigma_8(\Omegam/0.3)^{0.5} = 0.68^{+0.11}_{-0.15}$.  This result is
fully consistent with other literature weak lensing studies which
demonstrated that, despite the larger uncertainties, our approach was
able to keep all systematics under control.

Having tested the quality of our shear catalogue, the next step will be to
carry out detailed cosmological studies with different systematics
carefully accounted for. Furthermore, our results will allow us to
detect galaxy clusters over a broad redshift range, and constrain
their mass distribution from VOICE shear catalogue.


\section*{Acknowledgements}

We thank Jun Zhang for helpful comments on shear measurement.
L.P.F. acknowledges the support from NSFC grants 11673018, 11722326 \&
11333001, STCSM grant 16ZR1424800 \& 188014066 and SHNU grant DYL201603.
Z.H.F. acknowledges the support from NSFC grants 11333001 and
11653001. X.K.L. acknowledges the support from YNU Grant KC1710708 and
General Financial Grant from China Postdoctoral Science Foundation
with Grant No. 2016M591006.  Support for G.P. is provided by the
Ministry of Economy, Development, and Tourism's Millennium Science
Initiative through grant IC120009, awarded to The Millennium Institute
of Astrophysics, MAS. M.R. acknowledges the support from PRIN MIUR
2015 ``Cosmology and Fundamental Physics: illuminating the Dark
Universe with Euclid''. M.V. acknowledges support from the European
Commission Research Executive Agency (FP7-SPACE-2013-1 GA 607254), the
South African Department of Science and Technology (DST/CON 0134/2014)
and the Italian Ministry for Foreign Affairs and International
Cooperation (PGR GA ZA14GR02).


\bibliographystyle{mnras}

\begin{thebibliography}{}
\makeatletter
\relax
\def\mn@urlcharsother{\let\do\@makeother \do\$\do\&\do\#\do\^\do\_\do\%\do\~}
\def\mn@doi{\begingroup\mn@urlcharsother \@ifnextchar [ {\mn@doi@}
  {\mn@doi@[]}}
\def\mn@doi@[#1]#2{\def\@tempa{#1}\ifx\@tempa\@empty \href
  {http://dx.doi.org/#2} {doi:#2}\else \href {http://dx.doi.org/#2} {#1}\fi
  \endgroup}
\def\mn@eprint#1#2{\mn@eprint@#1:#2::\@nil}
\def\mn@eprint@arXiv#1{\href {http://arxiv.org/abs/#1} {{\tt arXiv:#1}}}
\def\mn@eprint@dblp#1{\href {http://dblp.uni-trier.de/rec/bibtex/#1.xml}
  {dblp:#1}}
\def\mn@eprint@#1:#2:#3:#4\@nil{\def\@tempa {#1}\def\@tempb {#2}\def\@tempc
  {#3}\ifx \@tempc \@empty \let \@tempc \@tempb \let \@tempb \@tempa \fi \ifx
  \@tempb \@empty \def\@tempb {arXiv}\fi \@ifundefined
  {mn@eprint@\@tempb}{\@tempb:\@tempc}{\expandafter \expandafter \csname
  mn@eprint@\@tempb\endcsname \expandafter{\@tempc}}}

\bibitem[\protect\citeauthoryear{{Aihara} et~al.,}{{Aihara}
  et~al.}{2018}]{2018PASJ...70S...4A}
{Aihara} H.,  et~al., 2018, \mn@doi [\pasj] {10.1093/pasj/psx066}, \href
  {http://adsabs.harvard.edu/abs/2018PASJ...70S...4A} {70, S4}

\bibitem[\protect\citeauthoryear{Anderson}{Anderson}{2003}]{Anderson2003}
Anderson T.~W.,  2003, An introduction to multivariate statistical analysis,
  3rd edn..
Wiley-Interscience

\bibitem[\protect\citeauthoryear{Barbary}{Barbary}{2016}]{Barbary2016}
Barbary K.,  2016, \mn@doi [The Journal of Open Source Software]
  {10.21105/joss.00058}, 1

\bibitem[\protect\citeauthoryear{{Bartelmann} \& {Schneider}}{{Bartelmann} \&
  {Schneider}}{2001}]{2001PhR...340..291B}
{Bartelmann} M.,  {Schneider} P.,  2001, \mn@doi [\physrep]
  {10.1016/S0370-1573(00)00082-X}, \href
  {http://ads.bao.ac.cn/abs/2001PhR...340..291B} {340, 291}

\bibitem[\protect\citeauthoryear{{Becker} et~al.,}{{Becker}
  et~al.}{2016}]{2016PhRvD..94b2002B}
{Becker} M.~R.,  et~al., 2016, \mn@doi [\prd] {10.1103/PhysRevD.94.022002},
  \href {http://adsabs.harvard.edu/abs/2016PhRvD..94b2002B} {94, 022002}

\bibitem[\protect\citeauthoryear{{Ben{\'{\i}}tez}}{{Ben{\'{\i}}tez}}{2011}]{Benitez2010}
{Ben{\'{\i}}tez} N.,  2011, {BPZ: Bayesian Photometric Redshift Code},
  Astrophysics Source Code Library (\mn@eprint {ascl} {1108.011})

\bibitem[\protect\citeauthoryear{{Ben{\'{\i}}tez} et~al.,}{{Ben{\'{\i}}tez}
  et~al.}{2004}]{Benitez2004}
{Ben{\'{\i}}tez} N.,  et~al., 2004, \mn@doi [\apjs] {10.1086/380120}, \href
  {http://adsabs.harvard.edu/abs/2004ApJS..150....1B} {150, 1}

\bibitem[\protect\citeauthoryear{{Benjamin} et~al.,}{{Benjamin}
  et~al.}{2013}]{CFHTLenS-2pt-tomo}
{Benjamin} J.,  et~al., 2013, \mn@doi [\mnras] {10.1093/mnras/stt276}, \href
  {http://adsabs.harvard.edu/abs/2013MNRAS.431.1547B} {431, 1547}

\bibitem[\protect\citeauthoryear{{Bertin}}{{Bertin}}{2011}]{2011ASPC..442..435B}
{Bertin} E.,  2011, in {Evans} I.~N.,  {Accomazzi} A.,  {Mink} D.~J.,   {Rots}
  A.~H.,  eds,  Astronomical Society of the Pacific Conference Series Vol. 442,
  Astronomical Data Analysis Software and Systems XX. p.~435

\bibitem[\protect\citeauthoryear{{Bertin} \& {Arnouts}}{{Bertin} \&
  {Arnouts}}{1996}]{Bertin1996}
{Bertin} E.,  {Arnouts} S.,  1996, \mn@doi [\aaps] {10.1051/aas:1996164}, \href
  {http://adsabs.harvard.edu/abs/1996A%26AS..117..393B} {117, 393}

\bibitem[\protect\citeauthoryear{{Botticella} et~al.,}{{Botticella}
  et~al.}{2017}]{2017A&A...598A..50B}
{Botticella} M.~T.,  et~al., 2017, \mn@doi [\aap]
  {10.1051/0004-6361/201629432}, \href
  {http://cdsads.u-strasbg.fr/abs/2017A%26A...598A..50B} {598, A50}

\bibitem[Capaccioli \& Schipani(2011)]{VST} Capaccioli, M., \& Schipani, P.\ 2011, The Messenger, 146, 2 

\bibitem[\protect\citeauthoryear{{Cappellaro} et~al.,}{{Cappellaro}
  et~al.}{2015}]{2015A&A...584A..62C}
{Cappellaro} E.,  et~al., 2015, \mn@doi [\aap] {10.1051/0004-6361/201526712},
  \href {http://adsabs.harvard.edu/abs/2015A%26A...584A..62C} {584, A62}

\bibitem[\protect\citeauthoryear{{Chang} et~al.,}{{Chang}
  et~al.}{2013}]{2013MNRAS.434.2121C}
{Chang} C.,  et~al., 2013, \mn@doi [\mnras] {10.1093/mnras/stt1156}, \href
  {http://adsabs.harvard.edu/abs/2013MNRAS.434.2121C} {434, 2121}

\bibitem[\protect\citeauthoryear{{Coleman}, {Wu}  \& {Weedman}}{{Coleman}
  et~al.}{1980}]{Coleman1980}
{Coleman} G.~D.,  {Wu} C.-C.,   {Weedman} D.~W.,  1980, \mn@doi [\apjs]
  {10.1086/190674}, \href {http://adsabs.harvard.edu/abs/1980ApJS...43..393C}
  {43, 393}

\bibitem[\protect\citeauthoryear{{Crittenden}, {Natarajan}, {Pen}  \&
  {Theuns}}{{Crittenden} et~al.}{2002}]{2002ApJ...568...20C}
{Crittenden} R.~G.,  {Natarajan} P.,  {Pen} U.-L.,   {Theuns} T.,  2002,
  \mn@doi [\apj] {10.1086/338838}, \href
  {http://adsabs.harvard.edu/abs/2002ApJ...568...20C} {568, 20}

\bibitem[\protect\citeauthoryear{{de Jong} et~al.,}{{de Jong}
  et~al.}{2015}]{2015A&A...582A..62D}
{de Jong} J.~T.~A.,  et~al., 2015, \mn@doi [\aap]
  {10.1051/0004-6361/201526601}, \href
  {http://adsabs.harvard.edu/abs/2015A%26A...582A..62D} {582, A62}


\bibitem[\protect\citeauthoryear{{de Jong} et~al.,}{{de Jong}
  et~al.}{2017}]{2017A&A...604A.134D}
{de Jong} J.~T.~A.,  et~al., 2017, \mn@doi [\aap]
  {10.1051/0004-6361/201730747}, \href
  {http://adsabs.harvard.edu/abs/2017A%26A...604A.134D} {604, A134}

  \bibitem[\protect\citeauthoryear{{De Cicco} et~al.,}{{De Cicco}
  et~al.}{2015}]{2015A&A...574A.112D}
{De Cicco} D.,  et~al., 2015, \mn@doi [\aap] {10.1051/0004-6361/201424906},
  \href {http://adsabs.harvard.edu/abs/2015A%26A...574A.112D} {574, A112}
    
    
\bibitem[\protect\citeauthoryear{{Falocco} et~al.,}{{Falocco}
  et~al.}{2015}]{2015A&A...579A.115F}
{Falocco} S.,  et~al., 2015, \mn@doi [\aap] {10.1051/0004-6361/201425111},
  \href {http://cdsads.u-strasbg.fr/abs/2015A%26A...579A.115F} {579, A115}

\bibitem[\protect\citeauthoryear{{Fu} \& {Fan}}{{Fu} \&
  {Fan}}{2014}]{2014RAA....14.1061F}
{Fu} L.-P.,  {Fan} Z.-H.,  2014, \mn@doi [Research in Astronomy and
  Astrophysics] {10.1088/1674-4527/14/9/002}, \href
  {http://ads.bao.ac.cn/abs/2014RAA....14.1061F} {14, 1061}

\bibitem[\protect\citeauthoryear{{Fu} et~al.,}{{Fu}
  et~al.}{2014}]{CFHTLenS-3pt}
{Fu} L.,  et~al., 2014, \mn@doi [\mnras] {10.1093/mnras/stu754}, \href
  {http://adsabs.harvard.edu/abs/2014MNRAS.441.2725F} {441, 2725}

\bibitem[\protect\citeauthoryear{{Gaia Collaboration} et~al.,}{{Gaia
  Collaboration} et~al.}{2016}]{2016A&A...595A...2G}
{Gaia Collaboration} et~al., 2016, \mn@doi [\aap]
  {10.1051/0004-6361/201629512}, \href
  {http://adsabs.harvard.edu/abs/2016A%26A...595A...2G} {595, A2}

\bibitem[\protect\citeauthoryear{{Grado}, {Capaccioli}, {Limatola}  \&
  {Getman}}{{Grado} et~al.}{2012}]{2012MSAIS..19..362G}
{Grado} A.,  {Capaccioli} M.,  {Limatola} L.,   {Getman} F.,  2012, Memorie
  della Societa Astronomica Italiana Supplementi, \href
  {http://adsabs.harvard.edu/abs/2012MSAIS..19..362G} {19, 362}

\bibitem[\protect\citeauthoryear{{Hartlap}, {Simon}  \& {Schneider}}{{Hartlap}
  et~al.}{2007}]{2007A&A...464..399H}
{Hartlap} J.,  {Simon} P.,   {Schneider} P.,  2007, \mn@doi [\aap]
  {10.1051/0004-6361:20066170}, \href
  {http://adsabs.harvard.edu/abs/2007A%26A...464..399H} {464, 399}

\bibitem[\protect\citeauthoryear{{Heymans} et~al.,}{{Heymans}
  et~al.}{2012a}]{CFHTLenS-sys}
{Heymans} C.,  et~al., 2012a, \mn@doi [\mnras]
  {10.1111/j.1365-2966.2012.21952.x}, \href
  {http://adsabs.harvard.edu/abs/2012MNRAS.427..146H} {427, 146}

\bibitem[\protect\citeauthoryear{{Heymans} et~al.,}{{Heymans}
  et~al.}{2012b}]{2012MNRAS.427..146H}
{Heymans} C.,  et~al., 2012b, \mn@doi [\mnras]
  {10.1111/j.1365-2966.2012.21952.x}, \href
  {http://adsabs.harvard.edu/abs/2012MNRAS.427..146H} {427, 146}

\bibitem[\protect\citeauthoryear{{Hildebrandt} et~al.,}{{Hildebrandt}
  et~al.}{2017}]{2017MNRAS.465.1454H}
{Hildebrandt} H.,  et~al., 2017, \mn@doi [\mnras] {10.1093/mnras/stw2805},
  \href {http://adsabs.harvard.edu/abs/2017MNRAS.465.1454H} {465, 1454}

\bibitem[\protect\citeauthoryear{{Hinshaw} et~al.,}{{Hinshaw}
  et~al.}{2013}]{2013ApJS..208...19H}
{Hinshaw} G.,  et~al., 2013, \mn@doi [\apjs] {10.1088/0067-0049/208/2/19},
  \href {http://adsabs.harvard.edu/abs/2013ApJS..208...19H} {208, 19}

\bibitem[\protect\citeauthoryear{{Huang}, {Radovich}, {Grado}, {Puddu},
  {Romano}, {Limatola}  \& {Fu}}{{Huang} et~al.}{2011}]{Huang2011}
{Huang} Z.,  {Radovich} M.,  {Grado} A.,  {Puddu} E.,  {Romano} A.,  {Limatola}
  L.,   {Fu} L.,  2011, \mn@doi [\aap] {10.1051/0004-6361/201015955}, \href
  {http://ads.bao.ac.cn/abs/2011A%26A...529A..93H} {529, A93}

\bibitem[\protect\citeauthoryear{{Jarvis} et~al.,}{{Jarvis}
  et~al.}{2013}]{Jarvis2013}
{Jarvis} M.~J.,  et~al., 2013, \mn@doi [\mnras] {10.1093/mnras/sts118}, \href
  {http://adsabs.harvard.edu/abs/2013MNRAS.428.1281J} {428, 1281}

\bibitem[\protect\citeauthoryear{{Jarvis} et~al.,}{{Jarvis}
  et~al.}{2016}]{2016MNRAS.460.2245J}
{Jarvis} M.,  et~al., 2016, \mn@doi [\mnras] {10.1093/mnras/stw990}, \href
  {http://adsabs.harvard.edu/abs/2016MNRAS.460.2245J} {460, 2245}

\bibitem[\protect\citeauthoryear{{Jee}, {Tyson}, {Schneider}, {Wittman},
  {Schmidt}  \& {Hilbert}}{{Jee} et~al.}{2013}]{2013ApJ...765...74J}
{Jee} M.~J.,  {Tyson} J.~A.,  {Schneider} M.~D.,  {Wittman} D.,  {Schmidt} S.,
   {Hilbert} S.,  2013, \mn@doi [\apj] {10.1088/0004-637X/765/1/74}, \href
  {http://adsabs.harvard.edu/abs/2013ApJ...765...74J} {765, 74}

\bibitem[\protect\citeauthoryear{{Jee}, {Tyson}, {Hilbert}, {Schneider},
  {Schmidt}  \& {Wittman}}{{Jee} et~al.}{2016}]{2016ApJ...824...77J}
{Jee} M.~J.,  {Tyson} J.~A.,  {Hilbert} S.,  {Schneider} M.~D.,  {Schmidt} S.,
   {Wittman} D.,  2016, \mn@doi [\apj] {10.3847/0004-637X/824/2/77}, \href
  {http://adsabs.harvard.edu/abs/2016ApJ...824...77J} {824, 77}

\bibitem[\protect\citeauthoryear{{Kaiser}}{{Kaiser}}{1992}]{1992ApJ...388..272K}
{Kaiser} N.,  1992, \apj, 388, 272

\bibitem[\protect\citeauthoryear{{Kilbinger}}{{Kilbinger}}{2015}]{2015RPPh...78h6901K}
{Kilbinger} M.,  2015, \mn@doi [Reports on Progress in Physics]
  {10.1088/0034-4885/78/8/086901}, \href
  {http://ads.bao.ac.cn/abs/2015RPPh...78h6901K} {78, 086901}

\bibitem[\protect\citeauthoryear{{Kilbinger} et~al.,}{{Kilbinger}
  et~al.}{2011}]{2011arXiv1101.0950K}
{Kilbinger} M.,  et~al., 2011, preprint, \href
  {http://adsabs.harvard.edu/abs/2011arXiv1101.0950K} {} (\mn@eprint {arXiv}
  {1101.0950})

\bibitem[\protect\citeauthoryear{{Kilbinger} et~al.,}{{Kilbinger}
  et~al.}{2013}]{2013MNRAS.430.2200K}
{Kilbinger} M.,  et~al., 2013, \mn@doi [\mnras] {10.1093/mnras/stt041}, \href
  {http://adsabs.harvard.edu/abs/2013MNRAS.430.2200K} {430, 2200}

\bibitem[\protect\citeauthoryear{{Kinney}, {Calzetti}, {Bohlin}, {McQuade},
  {Storchi-Bergmann}  \& {Schmitt}}{{Kinney} et~al.}{1996}]{Kinney1996}
{Kinney} A.~L.,  {Calzetti} D.,  {Bohlin} R.~C.,  {McQuade} K.,
  {Storchi-Bergmann} T.,   {Schmitt} H.~R.,  1996, \mn@doi [\apj]
  {10.1086/177583}, \href {http://adsabs.harvard.edu/abs/1996ApJ...467...38K}
  {467, 38}

\bibitem[\protect\citeauthoryear{{Kitching}, {Miller}, {Heymans}, {van
  Waerbeke}  \& {Heavens}}{{Kitching} et~al.}{2008}]{2008MNRAS.390..149K}
{Kitching} T.~D.,  {Miller} L.,  {Heymans} C.~E.,  {van Waerbeke} L.,
  {Heavens} A.~F.,  2008, \mn@doi [\mnras] {10.1111/j.1365-2966.2008.13628.x},
  \href {http://adsabs.harvard.edu/abs/2008MNRAS.390..149K} {390, 149}

\bibitem[\protect\citeauthoryear{{Kuijken} et~al.,}{{Kuijken}
  et~al.}{2015}]{2015MNRAS.454.3500K}
{Kuijken} K.,  et~al., 2015, \mn@doi [\mnras] {10.1093/mnras/stv2140}, \href
  {http://adsabs.harvard.edu/abs/2015MNRAS.454.3500K} {454, 3500}

\bibitem[\protect\citeauthoryear{{Liu}, {Wang}, {Pan}  \& {Fan}}{{Liu}
  et~al.}{2014}]{2014ApJ...784...31L}
{Liu} X.,  {Wang} Q.,  {Pan} C.,   {Fan} Z.,  2014, \mn@doi [\apj]
  {10.1088/0004-637X/784/1/31}, \href
  {http://adsabs.harvard.edu/abs/2014ApJ...784...31L} {784, 31}

\bibitem[\protect\citeauthoryear{{Liu} et~al.,}{{Liu}
  et~al.}{2015}]{2015MNRAS.450.2888L}
{Liu} X.,  et~al., 2015, \mn@doi [\mnras] {10.1093/mnras/stv784}, \href
  {http://adsabs.harvard.edu/abs/2015MNRAS.450.2888L} {450, 2888}

\bibitem[\protect\citeauthoryear{{Liu} et~al.,}{{Liu}
  et~al.}{2016}]{2016PhRvL.117e1101L}
{Liu} X.,  et~al., 2016, \mn@doi [Physical Review Letters]
  {10.1103/PhysRevLett.117.051101}, \href
  {http://adsabs.harvard.edu/abs/2016PhRvL.117e1101L} {117, 051101}


\bibitem[\protect\citeauthoryear{{Liu} et~al.,}{{Liu} et~al.}{2018}]{2017LiuDZ}
{Liu} D.,  et~al., 2018, \mn@doi [\mnras] {10.1093/mnras/sty1219}, \href
  {http://adsabs.harvard.edu/abs/2018MNRAS.tmp.1164L} {in press}

\bibitem[\protect\citeauthoryear{{Mandelbaum}}{{Mandelbaum}}{2017}]{2017arXiv171003235M}
{Mandelbaum} R.,  2017, preprint, \href
  {http://adsabs.harvard.edu/abs/2017arXiv171003235M} {} (\mn@eprint {arXiv}
  {1710.03235})

\bibitem[\protect\citeauthoryear{{Mandelbaum} et~al.,}{{Mandelbaum}
  et~al.}{2018}]{2018PASJ...70S..25M}
{Mandelbaum} R.,  et~al., 2018, \mn@doi [\pasj] {10.1093/pasj/psx130}, \href
  {http://adsabs.harvard.edu/abs/2018PASJ...70S..25M} {70, S25}

\bibitem[\protect\citeauthoryear{{Miller}, {Kitching}, {Heymans}, {Heavens}  \&
  {van Waerbeke}}{{Miller} et~al.}{2007}]{2007MNRAS.382..315M}
{Miller} L.,  {Kitching} T.~D.,  {Heymans} C.,  {Heavens} A.~F.,   {van
  Waerbeke} L.,  2007, \mn@doi [\mnras] {10.1111/j.1365-2966.2007.12363.x},
  \href {http://adsabs.harvard.edu/abs/2007MNRAS.382..315M} {382, 315}

\bibitem[\protect\citeauthoryear{{Miller} et~al.,}{{Miller}
  et~al.}{2013}]{2013MNRAS.429.2858M}
{Miller} L.,  et~al., 2013, \mn@doi [\mnras] {10.1093/mnras/sts454}, \href
  {http://adsabs.harvard.edu/abs/2013MNRAS.429.2858M} {429, 2858}

\bibitem[\protect\citeauthoryear{{Planck Collaboration} et~al.,}{{Planck
  Collaboration} et~al.}{2016}]{2016A&A...594A..13P}
{Planck Collaboration} et~al., 2016, \mn@doi [\aap]
  {10.1051/0004-6361/201525830}, \href
  {http://adsabs.harvard.edu/abs/2016A%26A...594A..13P} {594, A13}

\bibitem[\protect\citeauthoryear{{Rowe} et~al.,}{{Rowe}
  et~al.}{2015}]{2015A&C....10..121R}
{Rowe} B.~T.~P.,  et~al., 2015, \mn@doi [Astronomy and Computing]
  {10.1016/j.ascom.2015.02.002}, \href
  {http://adsabs.harvard.edu/abs/2015A%26C....10..121R} {10, 121}

\bibitem[\protect\citeauthoryear{{Schneider}, Van~Waerbeke  \&
  {Mellier}}{{Schneider} et~al.}{2002a}]{2002A&A...389..729S}
{Schneider} P.,  Van~Waerbeke L.,   {Mellier} Y.,  2002a, \aap, 389, 729

\bibitem[\protect\citeauthoryear{{Schneider}, {van Waerbeke}, {Kilbinger}  \&
  {Mellier}}{{Schneider} et~al.}{2002b}]{2002A&A...396....1S}
{Schneider} P.,  {van Waerbeke} L.,  {Kilbinger} M.,   {Mellier} Y.,  2002b,
  \mn@doi [\aap] {10.1051/0004-6361:20021341}, \href
  {http://adsabs.harvard.edu/abs/2002A%26A...396....1S} {396, 1}

\bibitem[\protect\citeauthoryear{{Schrabback} et~al.,}{{Schrabback}
  et~al.}{2010}]{2010A&A...516A..63S}
{Schrabback} T.,  et~al., 2010, \mn@doi [\aap] {10.1051/0004-6361/200913577},
  \href {http://adsabs.harvard.edu/abs/2010A%26A...516A..63S} {516, A63}

\bibitem[\protect\citeauthoryear{{Semboloni} et~al.,}{{Semboloni}
  et~al.}{2006}]{2006A&A...452...51S}
{Semboloni} E.,  et~al., 2006, \mn@doi [\aap] {10.1051/0004-6361:20054479},
  \href {http://adsabs.harvard.edu/abs/2006A%26A...452...51S} {452, 51}

\bibitem[\protect\citeauthoryear{{Skrutskie} et~al.,}{{Skrutskie}
  et~al.}{2006}]{2006AJ....131.1163S}
{Skrutskie} M.~F.,  et~al., 2006, \mn@doi [\aj] {10.1086/498708}, \href
  {http://adsabs.harvard.edu/abs/2006AJ....131.1163S} {131, 1163}

\bibitem[\protect\citeauthoryear{{Vaccari}}{{Vaccari}}{2015}]{2015fers.confE..27V}
  {Vaccari} M.,  2015, Proceedings of "The many facets of extragalactic radio
surveys: towards new scientific challenges" Conference, 20-23 October 2015,
Bologna, Italy, Proceedings of Science, 267, 27

\bibitem[\protect\citeauthoryear{{Vaccari} et~al.,}{{Vaccari}
  et~al.}{2010}]{2010A&A...518L..20V}
{Vaccari} M.,  et~al., 2010, \mn@doi [\aap] {10.1051/0004-6361/201014694},
  \href {http://adsabs.harvard.edu/abs/2010A%26A...518L..20V} {518, L20}

\bibitem[\protect\citeauthoryear{{Vaccari} et~al.,}{{Vaccari}
  et~al.}{2016}]{2016heas.confE..26V}
{Vaccari} M.,  et~al., 2016, Proceedings of the 4th Annual Conference on High
Energy Astrophysics in Southern Africa, 25-26 August 2016, Cape Town, South
Africa, Proceedings of Science, 275, 26


\makeatother
\end{thebibliography}

\begin{appendix}
\section{Photometric redshift using only optical bands }
\label{sec:zoptical}

\begin{figure}
    \resizebox{0.9\hsize}{!}{
      \includegraphics{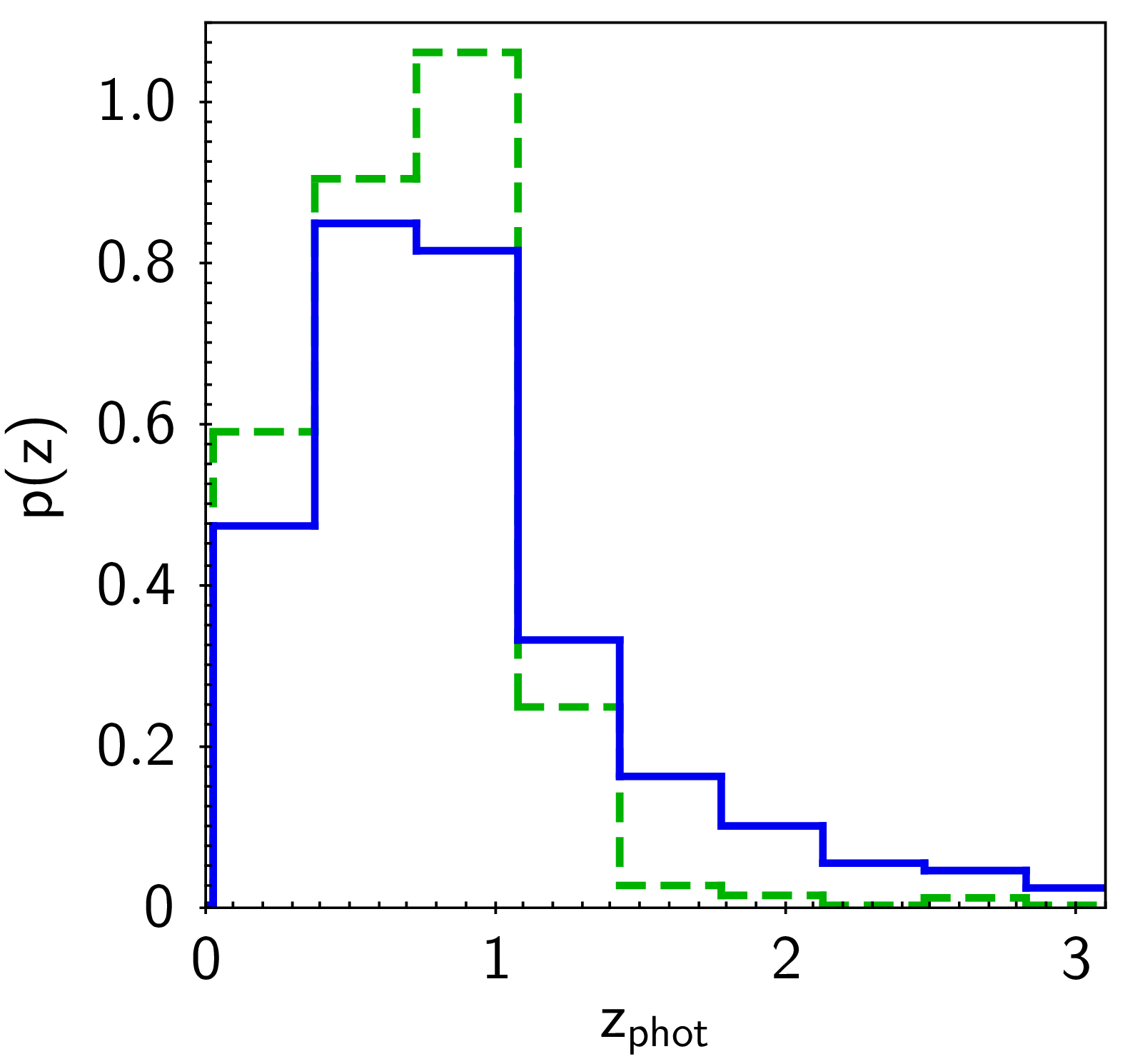}
      }
  \caption{The normalized  histogram of  photo-$z$ estimated using
    optical bands (4-band  photo-$z$, green dash line), optical and near-infared bands
  (8-band  photo-$z$, blue solid line) are shown, without considering
    the shear weight. }
   \label{fig:z4band}
\end{figure}

In order to show the improvement of photo-$z$ measurements by adding
near-infrared data, we estimate the photo-$z$ using VOICE optical
bands data (4-band photo-$z$) only.  We then match the 4-band
photo-$z$ catalog with the 8-band photo-$z$ for non-zero $\lensfit$
weight galaxies.  The redshift distribution histograms for the matched
galaxies are shown in Fig.~\ref{fig:z4band}. A significant difference is seen at
$z>1$ between the two photo-$z$ estimates. Without the near-infrared
data,  $\sim$15\%  of galaxies with 8-band $z>1$ are
assigned to lower redshifts.

  As in Sect.~\ref{sect:zdis}, we also compare the 4-band photo-$z$
  with the spec-$z$. The median value of $\delta z =
  ($photo-$z-$spec-$z)/(1+$spec-$z)$ and MAD values are $-0.010$ and
  $0.073$, respectively, which are $\sim$20\% larger than those of
  8-band photo-$z$ (see Table~\ref{tab:zcomp}).  We further separate
  galaxies into low-$z$ (4-band photo-$z <$ 0.83) and high-$z$
  (4-band photo-$z \geq$ 0.83) bins, and list median $\delta z$ and
  MAD values in Table~\ref{tab:zcomp}.  In comparison with the results
  of 8-band photo-$z$, about one-third of 8-band high-$z$ galaxies are
  shifted to the 4-band low-$z$ bin.  The offset $\delta z$ in
  high-$z$ bin is $\sim$3 times larger than that of 8-band photo-$z$.

\begin{table}
   \caption{ The number of spec-$z$ matched galaxies, their  median $\delta z$ and MAD
values are listed for all $z$,  low-$z$ and high-$z$ bins.   }

\begin{tabular}{|r|c|c|c|c|c|}

  \hline
          \hline
          & & & Ngal & $\delta z$ & MAD  \\
          \hline
          \multicolumn{2}{|c|}{\multirow{3}{*}{8-band photo-$z$}} &
          all  & 23638 & $-0.008$ &  0.060\\
          \multicolumn{2}{|c|}{} & low-$z$ & 19389 & $-0.012$  &          0.055 \\
          \multicolumn{2}{|c|}{} & high-$z$  & 4069 & 0.022 & 0.104\\ 
          \hline
          
            \multicolumn{2}{|c|}{\multirow{3}{*}{4-band photo-$z$}} &
          all  & 23638 & $-0.010$ & 0.073\\
          \multicolumn{2}{|c|}{} & low-$z$ & 20168 &  $-0.015$  & 0.067\\
          \multicolumn{2}{|c|}{} & high-$z$  & 3300 &$0.063$ & 0.160\\
         
        \hline 
        \end{tabular}
\label{tab:zcomp}
      \end{table}

\begin{figure}

 \resizebox{0.9\hsize}{!}{
    \includegraphics[bb =1 20 350 340 ]{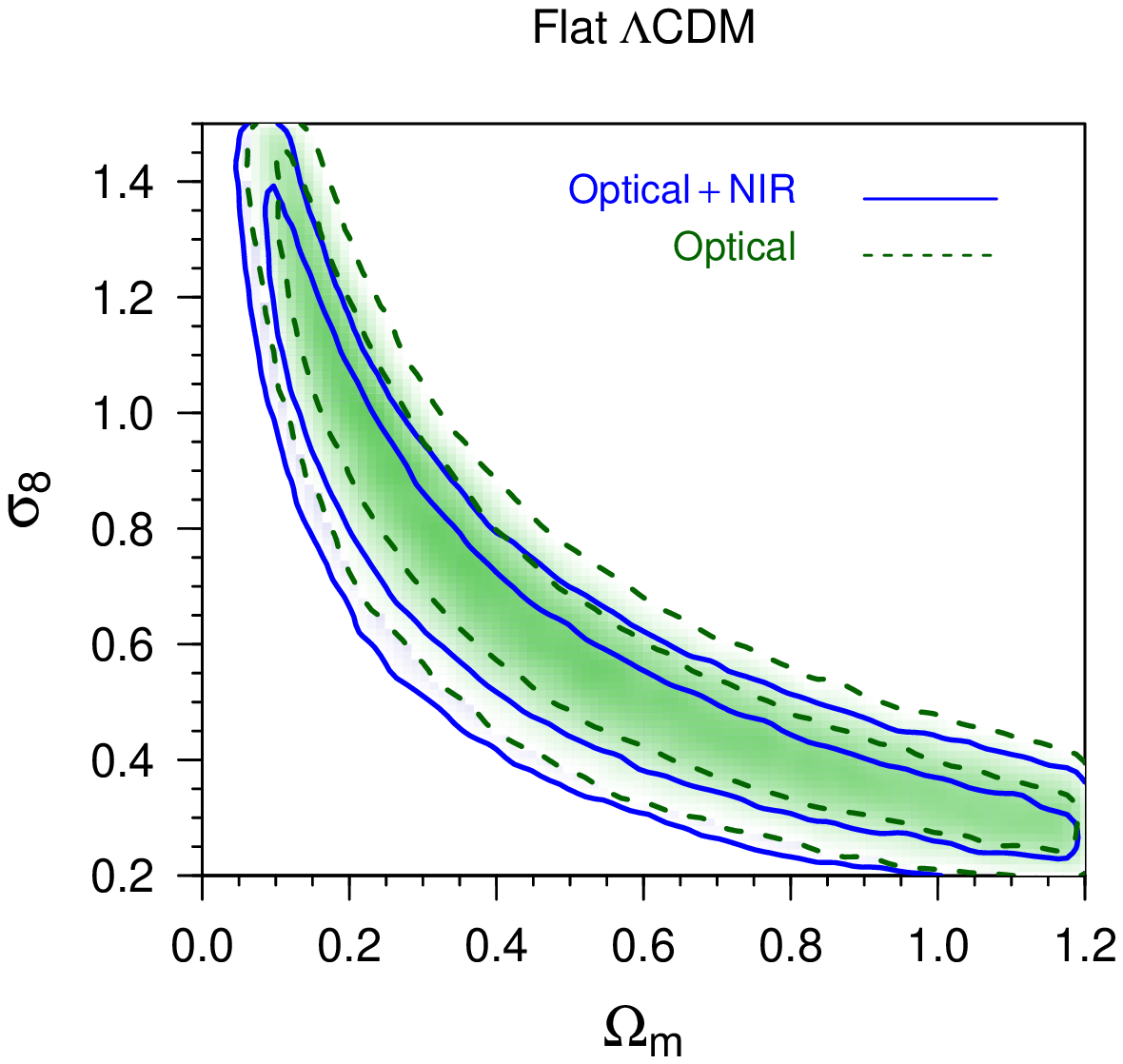}
 }
 \caption{Marginalized posterior density contours (68.3 per cent and  95.5 per
cent) for $\Omegam$ and $\sigma_8$ are constrained from 
     $\langle M_{\rm ap}^2\rangle$ in the case
    of flat $\Lambda$CDM. The blue contours are the constraints using
    8-band photo-$z$, while the green are the results using 4-band photo-$z$.}
   \label{fig:z4bandcos2}
\end{figure}

Fig.~\ref{fig:z4bandcos2} shows the cosmological constraints of
$\sigma_8$ and $\Omegam$ under the $\Lambda$CDM model using the 4-band
photo-$z$.  Compared to the constraints using 8-band photo-$z$, the
contours are shifted to the higher $\sigma_8$ and $\Omegam$ side.  The
$\Sigma_8 = \sigma_8(\Omegam/0.3)^{0.5}$ is shifted from
$0.68^{+0.11}_{-0.15}$ to $ 0.74^{+0.13}_{-0.16}$. 
Such a shift is in line with the fact that 15\% of the
high-$z$ galaxies in the 8-band photo-$z$ catalog are assigned to
low-$z$
bin.

\end{appendix}

\end{document}